\documentclass[12pt]{article}

\usepackage{amsmath}
\usepackage{amssymb}
\usepackage{amsfonts}
\usepackage{amsthm}
\usepackage{graphicx}
\usepackage{enumerate}
\usepackage{natbib}
\usepackage{url} 
\usepackage{color}
\usepackage{bm}
\usepackage{multirow}
\usepackage[hidelinks]{hyperref}
\usepackage{booktabs}
\usepackage{caption}
\usepackage{float}

\usepackage{array}
\usepackage{tikz}

\usetikzlibrary{arrows,shapes.arrows,shapes.geometric,
	shapes.multipart,backgrounds,decorations.pathmorphing,positioning,fit,automata}
\tikzset{
	>=stealth',
	truetwo/.style={
		rectangle,
		draw=black, very thick,
		text width=8.5em,
		minimum height=2em,
		text centered,
		fill=blue!20, opacity = 1.0},
	trueone/.style={
		rectangle,
		draw=black, very thick,
		text width=8.5em,
		minimum height=2em,
		text centered,
		fill=gray!50, opacity = 1.0},
	truethree/.style={
		rectangle,
		draw=black, very thick,
		text width=8.5em,
		minimum height=2em,
		text centered,
		fill=orange, opacity = 1.0},
	punkt/.style={
		rectangle,
		rounded corners,
		draw=black, very thick,
		text width=6.5em,
		minimum height=2em,
		text centered},
	est/.style={
		rectangle,
		draw=black, very thick,
		text centered},
	mytext/.style={
			rectangle,
			draw=white, very thick,
			text width=6.5em,
			minimum height=2em,
			text centered},
	shade/.style={
		rectangle,
		draw=black, very thick, fill=gray!50,
		text centered},
	weight/.style={
		circle,
		draw=black, very thick,
		text width=6.5em,
		minimum height=2em,
		text centered},
	pil/.style={
		->,
		thick,
		shorten <=2pt,
		shorten >=2pt,},
	double/.style={
		<->,
		thick,
		shorten <=2pt,
		shorten >=2pt,},
	dash/.style={
		dashed,
		thick,
		shorten <=2pt,
		shorten >=2pt,},
	dashdouble/.style={
		<->,
		dashed,
		thick,
		shorten <=2pt,
		shorten >=2pt,},
LargeBlock/.style={rectangle, draw, fill=blue!20, text width=10em, text badly centered, rounded corners},
decision/.style = {diamond, draw, fill=red!20, text badly centered,text width=1.2cm},
block/.style = {rectangle, draw, fill=blue!20, text width=5em, text badly centered, rounded corners, minimum height=1em},
ImgBlock/.style = {rectangle, draw},
line/.style = {draw, -latex'},
cloud/.style = {draw, ellipse,fill=red!20, minimum height=1em},
EmptyAnchor/.style = {circle,minimum height=1em}

}
\allowdisplaybreaks

\DeclareMathOperator*{\argmin}{arg\,min}

\newcommand{\blind}{0}

\theoremstyle{plain}
\newtheorem{theorem}{Theorem}
\newtheorem{lemma}{Lemma}

\newtheorem{assumption}{Assumption}

\theoremstyle{remark}

\newenvironment{tablenotes}{\begin{list}{}{\leftmargin=0pt\itemindent=0pt}\footnotesize}{\end{list}}
\providecommand{\phantomsection}{}
\makeatletter
\@ifundefined{theHtheorem}{}{}
\@ifundefined{theHlemma}{}{}
\makeatother
\renewcommand*{\thefootnote}{\fnsymbol{footnote}}

\def\spacingset#1{\renewcommand{\baselinestretch}%
{#1}\small\normalsize} \spacingset{1}

\definecolor{myblue}{RGB}{33,59,251} 
\definecolor{lightgreen}{RGB}{146,208,80} 
\definecolor{ZurichRed}{rgb}{1, 0, 0} 
\definecolor{ZurichGreen}{rgb}{.196,.804,.196} 
\definecolor{ZurichOrange}{rgb}{1,.648,0} 
	\definecolor{gold}{rgb}{0.83, 0.69, 0.22}

\newcommand{\ind}{\mbox{$\perp \kern-5.5pt \perp$}}

\begin{document}

\if1\blind
{
  \bigskip
  \bigskip
  \bigskip
  \begin{center}
    \spacingset{1.5}
    {\LARGE\bf Multi-Task Learning for High-Dimensional Regression with Many Weak Instruments}
  \end{center}
  \medskip
  \spacingset{1}
} \fi

\if0\blind
{
  \begin{center}
    \spacingset{1.5}
    {\LARGE\bf Multi-Task Learning for High-Dimensional Regression with Many Weak Instruments}\\ \bigskip \bigskip
    \spacingset{1}
    {\large
Di Zhang$^{1\ast}$, Xuanyu Li$^{2\ast}$, and Baoluo Sun$^{1\dagger}$\\ \bigskip
\normalsize
$^1$ Department of Statistics and Data Science, National University of Singapore\\
\smallskip
$^2$ School of Mathematical Sciences, University of Chinese Academy of Sciences
}
  \end{center}
  \begingroup
  \renewcommand{\thefootnote}{}
  \footnotetext{$^\ast$Co-first authors. $^\dagger$Address for correspondence: Department of Statistics and Data Science, S16/06-106, Science Drive 2,
National University of Singapore, Singapore 117546. Email: stasb@nus.edu.sg.}
  \endgroup
  \addtocounter{footnote}{-1}
} \fi

\bigskip

\begin{abstract}
Many weak instrumental variables (IVs) are routinely used in the health and social sciences to improve identification and inference of the treatment effect of interest, along with a broad collection of data on potential confounding factors in the hope that the IV assumptions hold within each data stratum. We propose a new debiased continuous-updating generalized method of moments estimator with multi-task learning of the IV propensity scores to simultaneously address the biases from a diverging number of weak IVs as well as first-step regularized estimation of nuisance regression functions in high-dimensional potential confounding factors.  We develop a new multi-task learning theory for generalized linear models under a general sub-Gaussian design to establish valid inference in the many weak IVs asymptotic regime under appropriate sparsity conditions. We evaluate the proposed method via extensive Monte Carlo studies and an empirical application to investigate the returns to education.
\end{abstract}

\noindent%
\vfill

\newpage
\spacingset{1.45} 

\section{Introduction}
\subsection{Background and motivation}
One of the main challenges with drawing causal inferences from observational data is the inability to categorically rule out  unmeasured confounding of the treatment-outcome relationship. The instrumental variable (IV) framework is a quasi-experimental approach to  identify and estimate treatment effects of interest under such settings, by leveraging one or more IVs to extract unconfounded variation in the treatment. When the IVs are not randomized by design,  the analyst often also considers a broad collection of data on potential confounding factors,   in the hope that the IV assumptions hold   within each data stratum. This paper focuses on estimation and inference on treatment effects with not only many measured potential confounding factors, but also many weak IVs. Concretely, we consider estimation and inference for $\beta_0$ under the canonical linear IV regression model,
\begin{equation}
    \begin{gathered}
{Y}_i=D_i \beta_0+X^{\prime}_i \theta_0+\epsilon_i,\quad  {E}(\epsilon_i \mid Z_i,X_i )=0,\quad i=1,...,n,\label{eq:pls}
 \end{gathered}
\end{equation}
where $Y_i$ is the outcome of interest, $D_i$ is an univariate endogenous treatment, $Z_i=(Z_{1i},...,Z_{mi})^\prime$ is an $m$-dimensional vector of IVs, and $X_i=(X_{1i},...,X_{pi})^\prime$ is a  $p$-dimensional vector of measured factors. Control of the potential confounding factors entails estimation of several unknown functions of the data through regression functions, such as $\ell_0(x)={E}(Y_i\mid X_i=x)$, $r_0(x)={E}(D_i\mid X_i=x)$, $\lambda_0(z,x)={E}(D_i\mid Z_i=z,X_i=x)$ and the $m$-dimensional IV propensity score $a_0(x)={E}(Z_i\mid X_i=x)$.

Over the past decade, as data dimensions increase, a large body of work has been devoted to addressing the general challenges in settings where $p$ is potentially larger than $n$ \citep{belloni2012sparse,https://doi.org/10.1111/rssb.12026, farrell2015robust,chernozhukov2015post,chernozhukov2015valid,Chernozhukov2018ddml,athey2018approximate,smucler2019unifying,bradic2019sparsity,bradic2019minimax, ning2020robust,10.1214/19-AOS1824,tan2020regularized, dukes2020doubly,hirshberg2021augmented, dukes2021inference, avagyan2022high,chernozhukov2022locally, tang2023ultra,liu2023root,zeng2024causal}, typically by positing sparse generalized linear models which are estimated with Lasso penalties \citep{tibshirani1996regression} so that a relatively low-dimensional subset adequately captures the confounding effects. For the linear IV regression model, \citet{chernozhukov2015post,chernozhukov2015valid} developed post-selection/regularization inference with both $m\gg n$ and $p\gg n$ based on moment functions of the partialled-out form $$m_i(\beta,\eta)=\left[{Y}_i-\ell(X_i;\gamma)-\{D_i-r(X_i;\xi)\}\beta\right]\{\lambda(Z_i,X_i;\psi)-r(X_i;\xi)\},\quad \eta=(\gamma^{\prime},\xi^{\prime},\psi^{\prime})^{\prime},$$  under sparse linear models $\ell_0(x)=x^{\prime}\gamma_0$, $r_0(x)=x^{\prime}\xi_0$, and $\lambda_0(z,x)=(z^{\prime}, x^{\prime})\psi_0$.  \citet{chernozhukov2015post,chernozhukov2015valid} showed that the average moment satisfies a form of local robustness (orthogonality) which mitigates the regularization bias from first-step Lasso estimation of $\eta$.  As the first-stage coefficient $\psi_0$ is sparse, a relatively small subset of strong IVs contributes to identification. Other works in this direction include  \citet{donald2001choosing, belloni2012sparse,carrasco2015regularized,chen2020mostly,liu2020deep,guo2022robustness} and \citet{scheidegger2025inference}.

The setup considered in this paper is different in that the IV strengths can be dense and individually weak, but the number of IVs $m$ grows with $n$, which leads to identification and asymptotic normality under many weak IVs asymptotics  as $n$ grows to infinity \citep{bekker1994alternative,Chao:2005aa,han2006gmm}; see \citet{mikusheva2022inference,mikusheva2024weak} for recent reviews and references therein. Under this regime, there has been growing interest in estimation and inference with many  potential confounding factors \citep{repec:pri:econom:2018-16,chao2023jackknife,mikusheva2024weak,matsushita5367195cross}. For example, the seminal study of  \cite{angrist1991does} on returns to education can incorporate rich functional forms and interactions of  measured baseline factors.  

\subsection{Our contribution}

Our work in this paper broadens the scope of inference with a diverging number of weak IVs by allowing for potentially high-dimensional baseline confounding factors, $p\gg n$. Specifically, our contributions are threefold:
\begin{enumerate}
\item \textit{Regularized estimation  with many weak IVs:}  We propose debiasing of the  continuous-updating  generalized method of moments (GMM) estimator of \cite{hansen1996finite},  with two key ingredients to address the biases from both first-step regularized estimation of nuisance parameters and many weak IVs, respectively: (a) The use of the alternative partialled-out $m$-dimensional orthogonal moment,
$$g_i(\beta,{\eta})=\left[{Y}_i-\ell(X_i;\gamma)-\{D_i-r(X_i;\xi)\}\beta\right]\{Z_i-a(X_i;\alpha)\},\quad {\eta}=(\gamma^{\prime},\xi^{\prime},\alpha^{\prime})^{\prime},$$
where $a_0(x)=\{h(x^{\prime}\alpha_{0,j}),...,h(x^{\prime}\alpha_{0,m})\}^{\prime}$ is an $m$-dimensional vector of IV propensity scores and $h$ is a (possibly nonlinear) link function. IV strength after partialling out the confounding factors is captured by the $m$-dimensional vector $$G=-{E}[\{D_i-r_0(X_i)\}\{Z_i-a_0(X_i)\}],$$ which can be dense with many small effects.  (b) Simultaneous minimization over the treatment effect parameter in the optimal weighting matrix of the continuous-updated objective function, so that the limiting objective function can be uniquely minimized at the true parameter value \citep{han2006gmm,Newey:2009aa}; in contrast the debiased GMM estimator of \cite{chernozhukov2022locally} may be vulnerable to bias in the many weak IVs asymptotic regime. We also adopt  sample splitting to avoid Donsker-type conditions on ${\eta}$, and it can be shown that consistency and asymptotic normality of the debiased continuous-updating GMM holds as long as 
 $$s_a \max \left( s_\ell  , s_r \right) \ll \frac{n}{m \log p \log pm},$$
 where $s_{a}$, $s_\ell$ and $s_r$ are the  sparsities of $a_0(x)$, $\ell_0(x)$ and $r_0(x)$ respectively. 

\item \textit{Multi-task learning of IV propensity scores:} A key challenge lies in learning a diverging number of IV propensity scores. To the best of our knowledge, our paper is the first to integrate multi-task learning (MTL) from the machine learning literature \citep{caruana1997multitask,lounici2009taking, Lounici_2011} in the high-dimensional IV regression framework. The main idea is that a relatively small, shared subset of measured factors suffices to control for the confounding effects of the IVs.  Under such structured sparsity assumptions, MTL can exploit the commonalities and borrow strength across the tasks of learning the $m$ IV propensity scores to produce more accurate estimates than single-task learning as $m$ diverges. Specifically, we establish consistency and asymptotic normality of the debiased continuous-updating GMM as long as 
 $$s_a \max \left( s_\ell  , s_r \right) \ll \frac{n}{(m+\log p)\log p},$$
which accommodates faster-growing dimensions compared to non-MTL. When $m \geq \log p$, MTL relaxes the sparsity requirement by a factor of $\log pm $.  Furthermore when $m \leq \log p$, MTL relaxes the sparsity requirement by an $m$ factor. This difference could be large with a large number of weak IVs. 

\item \textit{MTL theory for generalized linear models:}
Our work also makes advances in the MTL theory for generalized linear models. Many empirical applications involve binary or discrete IVs. In the multi-task linear regression setting, \cite{Lounici_2011} provide the convergence rate of the MTL estimator using group Lasso and demonstrate its advantages over the Lasso estimator. This result can be extended to the multi-task logistic regression model by directly applying existing results on the rate of the high-dimensional group Lasso estimator in logistic regression \citep{meier_group_2008, negahban_unified_2012,blazere_oracle_2014,van_de_geer_asymptotically_2014}. However, the existing results rely on a boundedness assumption on the potential confounding factors, limiting their practical applicability. We develop  new theory for the convergence rate of the multi-task logistic regression model. Our framework only requires a general sub-Gaussian design, which includes the boundedness design as a special case.
\end{enumerate} 

\subsection{Connections to existing work}
 The proposed debiased continuous-updating GMM framework with multi-task learning entails consistent estimation of the heteroscedastic optimal weighting matrix, which restricts the rate of growth $m$  relative to sample size $n$.  In particular, we need $m^3/n\rightarrow 0$ for asymptotic normality. Our work alleviates the impact of regularization bias on subsequent estimation and inference at the expense of a slower rate of growth of $m$, and thus 
 complements existing literature which allows $m$ to grow as fast as $n$, while the effects of confounding factors are typically partialled out without regularization, with $p$ growing at a rate slower than $n$  \citep{repec:pri:econom:2018-16,chao2023jackknife,mikusheva2024weak,matsushita5367195cross}. The type of asymptotics in our paper may be useful as an approximation in applications with high-dimensional baseline potential confounding factors, many weak IVs, and at the same time $n$ is still quite large relative to $m$. Besides the motivating study of  \cite{angrist1991does}, such applications are also common in the health sciences with advances in genotyping technology and large biobanks \citep{davies2015many}.

Sample splitting has been employed in many early works on semiparametric inference \citep{hasminskii1979nonparametric, bickel1982adaptive,pfanzagl1982lecture} and more recently \citet{ayyagari2010applications}, \citet{newey2018cross} and \citet{Chernozhukov2018ddml} in the context of partially linear models. The notion of Neyman orthogonality was introduced in  \citet{neyman1,neyman2} and has also played a key role in semiparametric learning theory \citep{andrews1994asymptotics, newey1994asymptotic, van2000asymptotic}. Methods grounded in semiparametric theory for regularized estimation in causal inference and missing or coarsened data problems abound, including double/debiased machine learning  \citep{chernozhukov2017double,Chernozhukov2018ddml} and targeted minimum loss-based estimation \citep{scharfstein1999adjusting, van2006targeted,van2011targeted,Zheng2011}, as discussed in \citet{diaz2020machine}. Our work represents an extension to the many weak IVs asymptotic regime, in which the number of moments can diverge with sample size while identification shrinks.

\citet{Newey:2009aa} established the limiting distribution for generalized empirical likelihood estimators including continuous-updating GMM under many weak moments asymptotics, but did not explicitly consider nuisance parameter estimation. \citet{ye2024genius} developed semiparametric continuous-updating GMM with a specific choice of first-step parametric and kernel estimators. Around the time of initial submission, we became aware of a preprint by \citet{wang2025gmm} on a similar continuous-updating GMM estimator with applications to structural mean models \citep{robins1994correcting,10.1214/14-STS493} and proximal causal inference \citep{miao2017invited,shi2020multiply, tchetgen2024introduction,cui2024semiparametric}. Our work introduces multi-task learning in the canonical high-dimensional linear IV regression model and thus complements theirs.

\subsection{Organization}
The remainder of the paper is organized as follows. We introduce the data structure and notation in Section \ref{sec:prelim}, before presenting the debiased continuous-updating GMM estimator with multi-task learning in Section \ref{sec:dcue}. We discuss how  multi-task learning can relax the sparsity requirements in Section \ref{sec:theo}, and evaluate the finite-sample performance of the debiased estimator through extensive Monte Carlo studies in Section \ref{sec:sim}. We revisit the \cite{angrist1991does} study in Section \ref{sec:app} and conclude in Section \ref{sec:discussion} with a brief discussion.

\section{Data and notation}
\label{sec:prelim}
Let $A^{\prime}$ denote the transpose of a vector or matrix $A$. The index set $\{1,...,n\}$ is denoted by $[n]$. Let $||v||=(\sum_{j=1}^q |v_j|^2)^{1/2}$  denote the Euclidean norm of a vector $v=(v_1,...,v_q)^{\prime}$, $||v||_1 = \sum_{j=1}^q |v_j|$ denote the $L_1$ norm of the vector $v$. For a symmetric matrix $A$, let $\lambda_{\min}(A)$ and $\lambda_{\max}(A)$ denote its smallest and largest eigenvalues, respectively. {For two sequences of real numbers $\{a_n\}_{n\in \mathbb{N}}$ and $\{b_n\}_{n\in \mathbb{N}}$, denote $a_n=o(b_n)$ if $a_n/b_n\rightarrow 0$ as $n\rightarrow\infty$, and $a_n=O(b_n)$ if $|a_n|\leq Cb_n$ for all $n$ and some $C>0$.} {In addition, for a sequence of random variables $\{L_n\}_{n\in \mathbb{N}}$,  denote $L_n=o_p(a_n)$ if for any $\epsilon>0$ ${P}(|L_n|/a_n \geq \epsilon)\rightarrow0$, as $n\rightarrow\infty$ and denote $L_n=O_p(a_n)$ if $\limsup_{n\rightarrow\infty}{P}(|L_n|/a_n \geq C)\rightarrow0$, as $C\rightarrow\infty$.}  Throughout the rest of the paper, $C$ is a positive generic constant which could vary in value depending on the context.

The data consists of i.i.d. observations $\{O_i=(Y_i,D_i,Z_i,X_i),i=1,...,n\}$. The dimensions of $Z_i=(Z_{1i},...,Z_{mi})^\prime$ and $X_i=(X_{1i},...,X_{pi})^\prime$ are $m$ and $p$, respectively. 
Throughout the paper, we refer to $\ell_0(x)= {E}(Y\mid X=x)=x'\gamma_0$, $r_0(x)= {E}(D\mid X=x)=x'\xi_0$ and $a_{0,j}(x)= {E}(Z_j\mid X=x)=h(x'\alpha_{0,j})$ as the outcome regression, treatment regression, and the $j$-th IV propensity score, respectively. Here $h$ is a known (possibly non-linear) link function. For ease of exposition and to simplify the proofs, we focus on logistic regression as a representative example. We refer to $a_0(x) = \{a_{0,1}(x), \dots, a_{0,m}(x)\}^\prime$ as the $m$-dimensional IV propensity score. The regression coefficients can be collected as $\eta_0=(\gamma^{\prime}_0,\xi^{\prime}_0,\alpha^{\prime}_0)^\prime$, where $\alpha_0=(\alpha^{\prime}_{0,1},\dots, \alpha^{\prime}_{0,m})^\prime$. In addition, we define the following quantities based on the $m$-dimensional function $g_i(\beta,{\eta})$:
\begin{align*}
g_i=g_i(\beta_0,\eta_0),\quad  {\Omega}= {E}({g}_i {g}_i^\prime) ,\quad  G_i(\eta)={\partial g_i(\beta, {{\eta}})}/{\partial \beta}, \quad G_i=G_i(\eta_0),\quad G={E}(G_i).
\end{align*}

\section{Multi-task learning of the IV propensity scores}
\label{sec:dcue}
In the empirical example of \cite{angrist1991does}, all the binary IVs are constructed from interactions between year-of-birth and quarter-of-birth. Therefore, if a given potential confounding factor is relevant for {\it any} of the IVs, it may also be expected to remain relevant for {\it all} of the IVs, e.g. demographic factors that correlate with birth-timing. This motivates the following multi-task sparse structure across the $m$ IV propensity scores. 
\begin{assumption}[Group sparsity]
\label{assum:glmgroupsparse}
   The regression coefficients $\alpha_0=(\alpha^{\prime}_{0,1},\dots, \alpha^{\prime}_{0,m})^\prime$ indexing the IV propensity score exhibit joint sparsity, that is, there exists a subset $S \subset [p]$ of cardinality $s_a$ that indexes all the nonzero components of $\alpha_{0,j}$, $j=1,...,m$.
\end{assumption}

We use cross-fitting, a form of sample splitting, to construct the continuous-updating debiased GMM estimator. Consider a $K$-fold random partition $\{I_k\}_{k=1}^K$ of the observation indices $[n]$ such that each fold $I_k$ contains $s= n/K$ samples. We assume that $s$ is an integer throughout for ease of presentation. Let $\hat{\eta}_k=(\hat{\gamma}_k,\hat{\xi}_k,\hat{\alpha}^{\prime}_k)^{\prime}$ denote the first-step regularized estimator of $\eta_0$ with MTL using only observations not in $I_k$, i.e. $\{O_i,i\notin  I_k\}$. Specifically, the multi-task generalized linear model estimator of $\alpha_0$ is the solution $\hat{\alpha}_k = (\hat\alpha^{\prime}_{k,1},\dots, \hat\alpha^{\prime}_{k,m})^\prime$ to the optimization problem,
\begin{align}
\label{eq:multitask}
     \underset{\alpha_1,...,\alpha_m} {\min}\left[\frac{1}{m(K-1)s}\sum_{j=1}^m \sum_{i\notin  I_k}\left\{- Z_{ji} X_i^T\alpha_j + \log (1+e^{X_i^T\alpha_j})\right\} + 2 \lambda_{1,n}\sum_{l=1}^p (\sum_{j=1}^m \alpha_{jl}^2)^{1/2} \right].
\end{align}
Since the group Lasso penalty \citep{Yuan_2005} selects the variables in a group in an "all-in" and "all-out" fashion, it ensures a measured confounding factor is either selected jointly across all IVs or excluded entirely. This structured regularization reflects the domain intuition of shared relevance across IVs, and enhances statistical efficiency by borrowing strength across tasks. Partially shared sparsity structure could also be leveraged in joint learning of the IV propensity scores \citep{behdin2025multi}. In addition, we use the Lasso estimator to estimate the parameters $\xi$ and $\gamma$ indexing the treatment and outcome regressions, respectively:
\begin{align}
\hat{\xi}_k &= \underset{\xi}{\min} \left\{\frac{1}{(K-1)s} \sum_{i\notin  I_k}^n(D_i- X_i ^T \xi )^2 + 2 \lambda_{2,n} ||\xi||_1 \right\}, \label{eq:lasso1} \\
\hat{\gamma}_k &= \underset{\gamma}{\min} \left\{ \frac{1}{(K-1)s}\sum_{i\notin  I_k}^n(Y_i -  X_i ^T \gamma )^2 + 2 \lambda_{3,n} ||\gamma||_1\right\}. \label{eq:lasso2}
\end{align}
The debiased continuous-updating GMM estimator is defined as 
\begin{equation}
\begin{aligned}
\hat{\beta}=\argmin_{\beta\in{\mathcal B}}\hat{Q}(\beta),\quad  \hat{Q}(\beta)=\hat{g}(\beta)^\prime  \hat{\Omega}^{-1}({\beta})  \hat{g}(\beta)/2,
\end{aligned}
\end{equation}
\noindent  where $\mathcal B$ is a compact set, and the empirical objective function is constructed using cross-fitted sample moments,
\begin{equation*}
\begin{aligned}
\hat{g}(\beta)=\frac{1}{n}\sum_{k=1}^K\sum_{i\in I_k}g_i(\beta,\hat{\eta}_k),\quad \hat{\Omega}(\beta)=\frac{1}{n}\sum_{k=1}^K\sum_{i\in I_k}g_i(\beta,\hat{\eta}_k)g^{\prime}_i(\beta,\hat{\eta}_k).
\end{aligned}
\end{equation*}
The moment function $g_i(\beta,\eta)$ yields average moments that are known to be orthogonal \citep{okui2012doubly,Chernozhukov2018ddml},
$$\partial E\{g_i(\beta_0,\eta)\}/\partial \eta\mid_{\eta=\eta_0}=0.$$
The cross-fitting scheme further mitigates bias arising from averaging over observations that are used to construct the first-step regularized estimators and information loss from sample-splitting,  without invoking Donsker-type conditions  on $\eta$. Although  orthogonal moments and cross-ﬁtting  are also used in the debiased GMM estimator
$$\tilde{\beta}=\argmin_{\beta\in \mathcal B}\tilde{Q}(\beta),\quad  \tilde{Q}(\beta)=\hat{g}(\beta)^\prime  \hat{W}  \hat{g}(\beta)/2,$$
where $\hat{W}$ is a positive semi-deﬁnite weighting matrix, it may be vulnerable to bias under many weak IVs asymptotics, as the signal about the treatment of interest captured in the average moments no longer dominates the noise of empirical moments. This is so even when a pre-computed optimal weight matrix $\hat{W}=\hat{\Omega}^{-1}({\check{\beta}})$ is used, where $\check{\beta}$ is some preliminary estimate. In contrast a key feature of the proposed debiased continuous-updating GMM is the simultaneous minimization over the treatment effect parameter $\beta$ in the moments and optimal weighting matrix, which eliminates $\beta$ from the noise term so that the objective function can be uniquely minimized at $\beta = \beta_0$ in the limit \citep{han2006gmm}.

\section{Theoretical properties}
\label{sec:theo}

\subsection{Convergence rate of multi-task learning}
\label{sec:rex-mul}
Let $S_r$ and $S_\ell$ be the supports of $\xi$ and $\gamma$, respectively. Denote $s_r = |S_r|$ and $s_\ell = |S_\ell|$ as the number of nonzero elements in  $\xi$ and $\gamma$. We make the following assumptions to facilitate our analysis, where $\hat{\alpha}$, $\hat{\xi}$, and $\hat{\gamma}$ denote the first-stage estimators in any given random split.

\begin{assumption}
\label{assum:randomdesign}
    $\{X_i\}_{i=1}^n$ are independent and identically distributed sub-Gaussian random vectors with parameter $c,$ that is, ${E} \exp\{v^{\prime} X_i\} \leq e^{||v||_2^2c^2/2}$ for all $v \in \mathbb{R}^p$. $\{Y_i - \ell_0(X_i)\}_{i=1}^n$ and $\{D_i - r_0(X_i)\}_{i=1}^n$ are independent and identically distributed sub-Gaussian random variables with parameter $c.$
\end{assumption}

\begin{assumption}
    \label{assum:parameterspace}
    Define $\Sigma_0 = E[X_i X_i^T] \in \mathbb{R}^{p\times p}$, we assume the parameter $(\alpha_0, \xi_0, \gamma_0, \Sigma_0)$ belong to the following parameter space indexed by the sparsity $s_a, s_r,$ and $s_\ell$,
\begin{align*}
    \Theta(s_a,s_r,s_\ell) = &\{(\alpha,\xi, \gamma, \Sigma)\mid \lvert\lvert\alpha_{k}\rvert\rvert_0 \leq s_a, \lvert\lvert\alpha_{k}\rvert\rvert \leq C, \lvert\lvert\xi \rvert\rvert_0 \leq s_r,\lvert\lvert\xi\rvert\rvert\leq C, \\
    &\lvert\lvert\gamma\rvert\rvert_0 \leq s_\ell,\lvert\lvert\gamma\rvert\rvert \leq C,   M^{-1} \leq \lambda_{\min} (\Sigma) \leq \lambda_{\max}(\Sigma) \leq M\},
\end{align*}
for constants $M>0$ and $C>0$ independent of $n$, $m$ and $p$.
\end{assumption}
\begin{assumption}
\label{assum:sparsity}
$s_a  = o\left(\min\left\{\frac{n}{m}, \frac{n}{\log p}\right\}\right), s_r = o\left(\frac{n}{\log p}\right), s_\ell = o\left(\frac{n}{\log p}\right)$.
\end{assumption}
The sub-Gaussian condition in Assumption \ref{assum:randomdesign} is a mild requirement that ensures the loss functions in the objective functions of \eqref{eq:multitask}--\eqref{eq:lasso2} exhibit local strong convexity properties \citep{negahban_unified_2012}. Assumption \ref{assum:parameterspace} prevents the parameter space from being ill-posed, a common constraint in high-dimensional regression models \citep{wainwright_sharp_2006, cai_statistical_2023}. Assumption \ref{assum:sparsity} is necessary for the consistency of the estimators $\hat{\alpha}$, $\hat{\xi}$, and $\hat{\gamma}$.

\begin{theorem}
\label{theorem:multirate}
    Under Assumptions \ref{assum:glmgroupsparse}--\ref{assum:sparsity}, there exist positive constants $c_1 $ such that with probability $1- p ^{-c_1}$, 
    \begin{align}
        \left(\sum_{j=1}^m ||\hat{\alpha}_{k,j} - \alpha_{0,j}||^2\right)^{1/2} & = O (\sqrt{\frac{s_a}{n}}(m+\log p)^{1/2}), \label{eq:multi-task-rate}\\
        || \hat{\xi}_k - \xi_0  || & = O( \sqrt\frac{s_r \log p }{n}), \\
         ||\hat{\gamma}_k - \gamma_0 ||& = O (\sqrt{\frac{s_\ell \log p}{n}}),
    \end{align}
    with the tuning parameter $\lambda_{1,n}, \lambda_{2,n}, \lambda_{3,n} $  chosen to be of order $ 
 \frac{1}{\sqrt{mn}}(1+\frac{\log p}{m})^{1/2},  \sqrt{\frac{\log p}{n}}$, and $   \sqrt{\frac{\log p}{n}}$.
\end{theorem}
Theorem \ref{theorem:multirate} establishes the rate of convergence for the MTL estimator under the sub-Gaussian random covariate design setting. The proof is novel and can be of independent interest. Based on a general framework for the consistency of high-dimensional M-estimation and results on random matrices, we remove the bounded design matrix assumption in the former literature on group Lasso binary outcome regression \citep{blazere_oracle_2014, koch_covariate_2018}.

\subsection{Relaxing the sparsity requirement by multi-task learning}
\label{sec:rex-mul2}
In this section, we describe the sparsity requirements so that multi-task learning yields sufficiently fast rates for key equicontinuity properties such as 
$$\sqrt{n}\hat{g}(\beta_0)=\frac{1}{\sqrt{n}}\sum_{i=1}^n g_i(\beta_0,\eta_0)+o_p(1),$$
to hold for asymptotic normality.
\begin{assumption}
\label{assum:sparsity2}
\begin{enumerate}[(i)]
    \item $s_a (m+\log p)\log p \max \left( s_\ell  , s_r \right) = o(n).$ 
    \item $s_a m (m +\log p) = o(n)$. 
    \end{enumerate}    
\end{assumption} 
Assumption \ref{assum:sparsity2} (i) is the multi-task learning version of rate doubly robustness \citep{smucler2019unifying,rotnitzky2021characterization,hines2022demystifying} which requires the product of the rates in Theorem \ref{theorem:multirate} to be sufficiently fast to ensure valid inference of the treatment effect parameter. It shows the possibility of admitting relatively non-sparse nuisance regression functions provided the remaining ones are sufficiently sparse. Assumption \ref{assum:sparsity2} (ii) is a minimal rate requirement to establish additional equicontinuity conditions such as for the empirical optimal weighting matrix which varies simultaneously in the treatment parameter of interest.  Our results also apply to the special case of a fixed, finite $m$ of strong IVs. In this case, both rate conditions in Assumption \ref{assum:sparsity2} are satisfied when $$s_a (\log p)^2 \max \left( s_\ell  , s_r \right) = o(n),$$ which agrees with the rate requirements for debiased machine learning  \citep{Chernozhukov2018ddml,chernozhukov2022locally}.

To illustrate the role of multi-task learning, suppose we ignore the group sparsity structure and use Lasso to estimate the $m$ IV propensity scores individually, which yields the non-MTL estimator $\tilde{\alpha}$. The lower bounds for the $L_2$ estimation errors of the  estimator $\tilde{\alpha}$  under certain conditions will be \citep{Lounici_2011},
\begin{align*}
     ||\tilde{\alpha} - \alpha_0|| \geq C  \sqrt{\frac{s_a m\log pm}{n}}.
\end{align*}
In addition, let $S_r$ and $S_\ell$ be the supports of $\xi$ and $\gamma$, respectively. Denote $s_r = |S_r|$ and $s_\ell = |S_\ell|$ as the number of nonzero elements in  $\xi$ and $\gamma$.  Thus the treatment and outcome regressions can be estimated at the rates $\sqrt{s_r \log p/n}$ and $\sqrt{s_\ell \log p/n}$, respectively. This necessitates the sparsity requirement
\begin{align}
\label{rate:Lasso}
    s_a m \log p \log pm \max \left( s_\ell  , s_r \right) = o(n).
\end{align}
To compare \eqref{rate:Lasso} and Assumption \ref{assum:sparsity2} (i), we consider two cases. When $m \geq \log p$, 
the sufficient rate requirement for the MTL estimator would relax the necessary rate requirement for the Lasso estimator by a factor of $\log pm $.  However, when $m \leq \log p$, the sufficient rate requirement for the MTL estimator would relax the necessary rate requirement for the Lasso estimator by a $m$ factor, which can make a large difference. The latter regime is likely when $p\gg n$ but $m$ grows more slowly than $n$, as in our case.

\subsection{Large sample inference}
\label{sec:theo2}

\setcounter{theorem}{1}

\begin{theorem}
\label{theorem::nor}
    Suppose there is a sequence of positive numbers $\mu_n$ such that $\frac{m}{\mu^{2}_n}$ is bounded, and $\mu_n\rightarrow \infty$, ${\mu^{-2}_n n G^\prime \Omega^{-1} G} \rightarrow \tau \in (0,\infty)$ as $n\rightarrow \infty$.  If Assumptions \ref{assum:glmgroupsparse}--\ref{assum:sparsity2} and additional regularity conditions in the Appendix hold,  then  as $n\rightarrow \infty$, $m^3/n\rightarrow 0$,
    \begin{equation*}
\mu_n(\hat{\beta}-\beta_0)\xrightarrow[]{d}N(0,V), \quad V=1/\tau+(\sigma/\tau)^2,
    \end{equation*}
    where $\sigma^2=\lim_{n\rightarrow \infty}\mu^{-2}_n {E}(U^{\prime}_i \Omega^{-1}U_i)$ and $U_i=G_i-G-\{\Omega^{-1}{E}(g_i G^{\prime}_i)\}^{\prime} g_i$ is the population residual from least squares regression of $G_i-G$ on $g_i$.
\end{theorem}
The first part of Theorem \ref{theorem::nor} characterizes the growth rate of $n G^\prime \Omega^{-1} G$, which may be viewed as an analog of the concentration parameter in weak IV literature \citep{stock2002survey}. It subsumes the classical GMM asymptotics with fixed $m$, in which case we can choose the sequence $\mu_n=\sqrt{n}$. More generally, $m$ can diverge with sample size $n$ while individual IV strengths shrink to achieve identification. For example, suppose $\Omega=\sigma^2{I}$ and $G=C n^{-1/2}(1,...,1)^{\prime}$ shrinking at $\sqrt{n}$-rate for some $C>0$ \citep{Staiger:1997aa}, in which case ${n} G^\prime \Omega^{-1} G=\left({C}/\sigma\right)^2 m$. Thus, if $m$ is fixed, then  the concentration parameter no longer diverges and identification fails. However, when $m$ diverges,  the first part of Theorem \ref{theorem::nor} is satisfied with $\tau=\left(C/\sigma\right)^2$ if we choose $\mu_n=\sqrt{m}$, and thus $\mu_n$ grows as fast as $\sqrt{m}$ \citep{chao2005consistent}. 

Wald-type inference can proceed in the usual way. An estimator of the asymptotic variance of $\hat{\beta}$ is $\hat{V}/n$, where
\begin{align*}
    \hat{V}&=\frac{\hat{D}^\prime\hat{\Omega}^{-1}\hat{D}}{\hat{H}^{2}},\quad {\hat{H}}=\frac{\partial^2\hat{Q}(\hat{\beta})}{\partial\beta^2},\quad \hat{\Omega}=\hat{\Omega}(\hat{\beta}), \\
    \hat{D}&=\hat{D}(\hat{\beta})= \frac{\partial \hat{g}(\hat{\beta})}{\partial \beta}-\frac{1}{n}\sum_{k=1}^K\sum_{i\in I_k}\{G_i(\hat{\eta}_k)g_i(\hat{\beta},\hat{\eta}_k)^\prime\}\hat{\Omega}^{-1}\hat{g}(\hat{\beta}).
\end{align*}
\noindent To test the null hypothesis $H_0: \beta=\beta_0,$  the Wald statistic $\hat{T}=n(\hat{\beta}-\beta_0)^2/\hat{V}\overset{d}{\rightarrow}\chi^2(1)$. We can also construct other identification robust statistics such as the  Lagrange multiplier statistic for generalized
empirical likelihood methods \citep{kleibergen2005testing, guggenberger2005generalized},
\begin{equation*}
    \hat{K}=n\frac{\partial \hat{Q}({\beta}_0)}{\partial{\beta}}\left\{\hat{D}(\beta_0)^{\prime}\hat{\Omega}({\beta}_0)\hat{D}(\beta_0)\right\}^{-1}\frac{\partial \hat{Q}({\beta}_0)}{\partial{\beta}},
\end{equation*}
which  does not require computation of the estimate $\hat{\beta}$ and is asymptotically equivalent to $\hat{T}$ under the many weak IV asymptotics \citep{Newey:2009aa}. Another important  direction concerns testing of overidentifying restrictions, which can be used to assess whether the model-implied moment restrictions are compatible with the data.  A Sargan–Hansen test  \citep{sargan1958estimation, Hansen:1982aa} can be constructed  based on the debiased $J$-statistic $2n\hat{Q}(\hat{\beta})$. The theoretical properties of the tests are summarized as follows.
\begin{theorem}
    \label{theorem:lsinfer}
    Under the same conditions in Theorem \ref{theorem::nor}, $\hat{V}$ is consistent, i.e. $\mu_n^2\hat{V}/n\overset{p}{\rightarrow}V$. Furthermore, as $n\rightarrow \infty$, (i) $\hat{T}\overset{d}{\rightarrow}\chi^2(1)$, (ii) $\hat{K}\overset{d}{\rightarrow}\hat T+o_p(1)$, and (iii) $P\left(2n\hat{Q}(\hat\beta)\geq q^{m-1}_{a} \right)\rightarrow a$, where $q^{m-1}_{a}$ is the $(1-a)$-th quantile of a $\chi^2(m-1)$ distribution.
\end{theorem}

\section{Monte Carlo Studies}
\label{sec:sim}

\subsection{Data generating mechanism}
We perform Monte Carlo experiments to investigate the finite-sample performance of the proposed debiased continuous-updating GMM estimator (CUE) under  a study design which follows that in \cite{Newey:2009aa} except for the presence of covariates. Let $\bar{Z}_i=Z_i-E(Z_i|X_i)$, $\bar{D}_i=D_i-E(D_i|X_i)$, and $\bar{Y}_i=Y_i-E(Y_i|X_i)$ for $i=1,...,n$.  The data generating process is given by
\begin{align*}
    &{X}_i=(X_{1i},...,X_{100i})^\prime\sim N(0,I),\quad  {Z}_i=(Z_{1i},...,Z_{mi})^\prime,\quad P(Z_{ji} = 1)=h({X}^\prime_i {\gamma}), \quad j=1,...,m,\\
        &E(D_i|X_i) = X_i^\prime{\lambda}_D,   \quad  \bar{D}_i=\bar{Z}^\prime_i {\pi}+\nu_i,\\  
    & E(Y_i|X_i) = {X}^\prime_i {\lambda}_Y, \quad
    \bar Y_i=\bar D_i\beta_0+\rho\nu_i+\sqrt{1-\rho^2}\varepsilon_i, \\
    &\epsilon_i\sim N(0,1),\quad \nu_i\sim N(0,1),\quad \varepsilon_i \sim N(0,1),\quad {\pi}=\left(\sqrt{\frac{\textup{CP}}{nm}},...\sqrt{\frac{\textup{CP}}{nm}}\right)^\prime,
\end{align*}
where ${\gamma}$, ${\lambda}_D$ and ${\lambda}_Y$ are all $100$-dimensional coefficient vectors with $0.2$ in the first $s$ entries and 0 in the remaining entries. To investigate the impact of sparsity assumption on the estimator, we further consider two sparsity scenarios: $s = 5$ representing the ultra-sparse scenario, and $s=10$ representing the mildly sparse scenario. The correlation coefficient between the structural and reduced form errors is $\rho=0.3$, the number of IVs is set to $m=15$ or 30, the concentration parameter which encodes IV strength, is set to $\textup{CP}=60$ or 120, and the sample size is $n=500$ and 1000. The parameter of interest is $\beta_0=0$. We implement multi-task debiased-CUE and single-task debiased-CUE by using the R packages \texttt{RMTL} \citep{cao_rmtl_2019} and \texttt{glmnet}, respectively. The following estimators are also included for comparison:
\begin{itemize}
\item[(i)] The conventional TSLS estimator based on linear models (in $X_i$) for the first-step regression functions.
\item[(ii)] The two-step debiased-GMM with identity weighting matrix, and multi-task GLM first-step estimation.
\item[(iii)] The jackknife IV estimator with many exogenous control variables and weak moments \citep{chao2023jackknife}.
\item[(iv)] As performance benchmarks, the oracle-CUE and two-step oracle-GMM with access to the true functional forms of the first-step regression functions (and thus no cross-fitting is involved).
\end{itemize}

\subsection{Monte Carlo results}

Tables \ref{tab:sim1} and \ref{tab:sim2} summarize the Monte Carlo results for the first scenario with $p=100$ covariates and $s = 5$ based on $1000$ replications with $n=500$ and 1000, respectively. Tables \ref{tab:sim3} and \ref{tab:sim4} summarize the Monte Carlo results for the second scenario with $p=100$ covariates and $s = 10$ based on $1000$ replications with $n=500$ and 1000, respectively. The bias of two-step oracle-GMM increases with the number of IVs and a decreasing concentration parameter value, which agrees with the Monte Carlo results in \cite{Newey:2009aa}. Our Monte Carlo studies indicate similar results for the bias of TSLS and two-step debiased-GMM in the presence of first-step estimation of the regression functions. Although debiased-CUE is substantially more dispersed, it yields noticeably smaller absolute biases and better coverage than TSLS and debiased-GMM. In particular, debiased-CUE shows negligible bias and nominal coverage when $n=1000$. To further compare the empirical performance of multi-task debiased-CUE and single-task debiased-CUE, we observe that both methods yield similar results under low sparsity conditions ($s=5$), while multi-task debiased-CUE outperforms the single-task debiased-CUE when sparsity increases ($s=10$), which is consistent with theoretical expectations. The multi-task debiased-CUE shows greater robustness to the sparsity level than the single-task counterpart. 
 \begin{table}[H]
    \caption{Monte Carlo results for $s = 5$ with $p=100$ covariates ($n=500$).}
    \label{tab:sim1}
    \centering
    \resizebox{.68\textwidth}{!}{\begin{tabular}{lcccccccc}
        \toprule
        Method & CP & $m$& $|\text{mean bias}|$ & $|\text{median bias}|$ & $\sqrt{\text{Var}}$ & $\text{ E} \sqrt{\text{Var}}$& Cov95 \\
        \midrule
        \multirow{4}{*}{multi-task-debiased-CUE}   & 60  &  15  &  .013  &  .089  &  1.263  &  1.576  &  0.947   \\
          &  120  &  15  &  .008  &  .043  &  .604  &  .527  &  0.940      \\
           &  60  &  30  &  .087  &  .133  &  1.713  &  1.978  &  0.895   \\
            &120  &  30  &  .064  &  .066  &  1.082  &  1.167  &  0.924\\
           \midrule
                \multirow{4}{*}{single-task-debiased-CUE}   & 60  &  15  &  .078  &  .144  &  1.226  &  1.277  &  0.947   \\
            &  120  &  15  &  .039  &  .065  &  .774  &  .763  &  0.940       \\  
            &  60  &  30  &  .149  &  .187  &  1.905  &  2.379  &  0.885     \\
            &   120  &  30  &  .109  &  .093  &  1.392  &  1.715  &  0.921     \\ 
        \midrule
                \multirow{4}{*}{TSLS}   &   60  &  15  &  .165  &  .170  &  .204  &  .201  &  0.844  \\
                &  120  &  15  &  .113  &  .118  &  .173  &  .168  &  0.871  \\
          &  60  &  30  &  .221  &  .217  &  .154  &  .154  &  0.689    \\
         & 120  &  30  &  .165  &  .160  &  .139  &  .137  &  0.759 \\
                 \midrule
        \multirow{4}{*}{debiased-GMM}   &   60  &  15  &  .188  &  .194  &  .196  &  .195  &  0.801 \\
              &    120  &  15  &  .142  &  .150  &  .166  &  .164  &  0.832 \\
            &   60  &  30  &  .246  &  .245  &  .152  &  .155  &  0.640 \\
            & 120  &  30  &  .201  &  .200  &  .133  &  .139  &  0.698     \\ 
        \midrule
        \multirow{4}{*}{oracle-CUE}   & 60  &  15  &  .066  &  .000  &  .872  &  .898  &  0.949\\
              & 120  &  15  &  .033  &  .007  &  .455  &  .380  &  0.950       \\
          & 60  &  30  &  .052  &  .040  &  1.401  &  1.444  &  0.924   \\ 
          & 120  &  30  &  .042  &  .003  &  .614  &  .603  &  0.945   \\
           \midrule
        \multirow{4}{*}{oracle-GMM}   & 60  &  15  &  .143  &  .149  &  .190  &  .189  &  0.850   \\
            &       120  &  15  &  .101  &  .104  &  .158  &  .155  &  0.873   \\
             &  60  &  30  &  .204  &  .203  &  .146  &  .150  &  0.705    \\
           & 120  &  30  &  .147  &  .148  &  .131  &  .130  &  0.778     \\
           \midrule
        \multirow{4}{*}{Jackknife}    &  60  &  15  &  .301  &  .024  &  20.683  &  372.439  &  0.969   \\
           &  120  &  15  &  .036  &  .014  &  1.709  &  3.061  &  0.947  \\
            &   60  &  30  &  .010  &  .112  &  7.393  &  100.072  &  0.959     \\
           &  120  &  30  &  1.117  &  .009  &  28.444  &  661.808  &  0.955   \\
        \bottomrule
    \end{tabular}}
           \begin{tablenotes}
      \item  {\noindent\footnotesize Note: $|\text{mean bias}|$,  $|\text{median bias}|$ and $\sqrt{\text{Var}}$ are the Monte Carlo mean absolute bias, median absolute bias  and standard deviation of the point estimates,  $\text{ E} \sqrt{\text{Var}}$ is the
mean of standard deviation estimates and Cov95 is the coverage proportion of
the 95\% Wald confidence intervals, based on 1000 repeated simulations.  
}
    \end{tablenotes}
\end{table}
 \begin{table}[H]
    \caption{Monte Carlo results for $s = 5$ with $p=100$ covariates ($n=1000$).}
    \label{tab:sim2}
    \centering
    \resizebox{.7\textwidth}{!}{\begin{tabular}{lcccccccc}
        \toprule
        Method & CP & $m$& $|\text{mean bias}|$ & $|\text{median bias}|$ & $\sqrt{\text{Var}}$ & $\text{ E} \sqrt{\text{Var}}$& Cov95 \\
        \midrule
        \multirow{4}{*}{multi-task-debiased-CUE}   &  60  &  15  &  .061  &  .073  &  .875  &  .976  &  0.946  \\
            & 120  &  15  &  .030  &  .061  &  .393  &  .344  &  0.961  \\
           & 60  &  30  &  .017  &  .080  &  1.183  &  1.341  &  0.927 \\
           &  120  &  30  &  .036  &  .047  &  .764  &  .898  &  0.934 \\
           \midrule
                \multirow{4}{*}{single-task-debiased-CUE}   &    60  &  15  &  .055  &  .105  &  1.062  &  1.306  &  0.938 \\
            &     120  &  15  &  .048  &  .082  &  .383  &  .355  &  0.955  \\  
            &   60  &  30  &  .063  &  .132  &  1.256  &  1.646  &  0.920  \\
            &  120  &  30  &  .091  &  .102  &  .824  &  .848  &  0.930        \\ 
        \midrule
                \multirow{4}{*}{TSLS}   
                &  60  &  15  &  .162  &  .165  &  .182  &  .191  &  0.857 \\
          &    120  &  15  &  .105  &  .114  &  .157  &  .160  &  0.887   \\
         &  60  &  30  &  .205  &  .202  &  .151  &  .150  &  0.693 \\
             &  120  &  30  &  .161  &  .158  &  .133  &  .134  &  0.747\\
             \midrule
        \multirow{4}{*}{debiased-GMM}   &   60  &  15  &  .186  &  .187  &  .174  &  .186  &  0.829 \\
              &   120  &  15  &  .133  &  .142  &  .150  &  .157  &  0.864  \\
            &   60  &  30  &  .220  &  .218  &  .147  &  .146  &  0.647  \\
            &   120  &  30  &  .181  &  .180  &  .130  &  .130  &  0.686   \\ 
        \midrule
        \multirow{4}{*}{oracle-CUE}   &  60  &  15  &  .003  &  .019  &  .791  &  .805  &  0.948   \\
              &  120  &  15  &  .025  &  .003  &  .306  &  .305  &  0.968    \\
          &  60  &  30  &  .040  &  .014  &  1.038  &  1.176  &  0.945 \\ 
          &   120  &  30  &  .049  &  .011  &  .713  &  .765  &  0.956 \\
           \midrule
        \multirow{4}{*}{oracle-GMM}   &  60  &  15  &  .154  &  .156  &  .180  &  .186  &  0.860  \\
            &    120  &  15  &  .096  &  .105  &  .153  &  .154  &  0.901     \\
             &    60  &  30  &  .197  &  .188  &  .145  &  .148  &  0.704  \\
           &  120  &  30  &  .150  &  .149  &  .130  &  .130  &  0.764    \\
           \midrule
        \multirow{4}{*}{Jackknife}    &     60  &  15  &  .061  &  .017  &  3.546  &  15.533  &  0.973  \\
           &   120  &  15  &  .110  &  .008  &  1.319  &  2.553  &  0.968 \\
            &    60  &  30  &  .127  &  .030  &  7.830  &  67.142  &  0.973   \\
           &   120  &  30  &  .105  &  .015  &  2.719  &  9.306  &  0.966  \\
        \bottomrule
    \end{tabular}}
           \begin{tablenotes}
      \item  {\noindent\footnotesize Note: See the footnote of Table \ref{tab:sim1}.

}
    \end{tablenotes}
\end{table}
 \begin{table}[H]
    \caption{Monte Carlo results for $s = 10$ with $p=100$ covariates ($n=500$).}
    \label{tab:sim3}
    \centering
    \resizebox{.7\textwidth}{!}{\begin{tabular}{lcccccccc}
        \toprule
        Method & CP & $m$& $|\text{mean bias}|$ & $|\text{median bias}|$ & $\sqrt{\text{Var}}$ & $\text{ E} \sqrt{\text{Var}}$& Cov95 \\
        \midrule
        \multirow{4}{*}{multi-task-debiased-CUE}   & 60  &  15  &  .157  &  .117  &  1.345  &  1.536  &  0.929  \\
           &  120  &  15  &  .046  &  .069  &  .616  &  .536  &  0.935   \\
           &  60  &  30  &  .114  &  .173  &  1.533  &  1.826  &  0.905  \\
           &   120  &  30  &  .117  &  .126  &  1.234  &  1.227  &  0.908\\
           \midrule
                \multirow{4}{*}{single-task-debiased-CUE}   & 60  &  15  &  .260  &  .202  &  1.498  &  1.932  &  0.925  \\
            &  120  &  15  &  .113  &  .124  &  .906  &  1.174  &  0.938    \\  
            &  60  &  30  &  .311  &  .283  &  1.838  &  2.615  &  0.895 \\
            &    120  &  30  &  .195  &  .228  &  1.428  &  1.823  &  0.909  \\ 
        \midrule
                \multirow{4}{*}{TSLS}   &  60  &  15  &  .175  &  .171  &  .206  &  .203  &  0.850  \\
                &  120  &  15  &  .115  &  .120  &  .171  &  .171  &  0.878  \\
          &  60  &  30  &  .223  &  .220  &  .154  &  .155  &  0.666  \\
         & 120  &  30  &  .172  &  .174  &  .140  &  .139  &  0.748 \\
                 \midrule
        \multirow{4}{*}{debiased-GMM}   &   60  &  15  &  .214  &  .211  &  .195  &  .191  &  0.781  \\
              & 120  &  15  &  .159  &  .165  &  .163  &  .162  &  0.798  \\
            &  60  &  30  &  .262  &  .263  &  .150  &  .152  &  0.573 \\
            &   120  &  30  &  .224  &  .220  &  .136  &  .138  &  0.620 \\ 
        \midrule
        \multirow{4}{*}{oracle-CUE}   & 60  &  15  &  .067  &  .013  &  1.130  &  1.129  &  0.955  \\
              &  120  &  15  &  .024  &  .001  &  .553  &  .445  &  0.962   \\
          &  60  &  30  &  .069  &  .000  &  1.337  &  1.468  &  0.933 \\ 
          &  120  &  30  &  .048  &  .010  &  .871  &  .719  &  0.947 \\
           \midrule
        \multirow{4}{*}{oracle-GMM}   &  60  &  15  &  .154  &  .160  &  .187  &  .191  &  0.869  \\
            &    120  &  15  &  .099  &  .100  &  .155  &  .158  &  0.896   \\
             &  60  &  30  &  .203  &  .206  &  .150  &  .150  &  0.704  \\
           &  120  &  30  &  .154  &  .158  &  .131  &  .131  &  0.761  \\
           \midrule
        \multirow{4}{*}{Jackknife}    &  60  &  15  &  .073  &  .051  &  5.909  &  26.996  &  0.968  \\
           &   120  &  15  &  .194  &  .034  &  1.931  &  5.770  &  0.961 \\
            &   60  &  30  &  1.997  &  .115  &  66.409  &  7080.591  &  0.968   \\
           &  120  &  30  &  .121  &  .002  &  1.569  &  3.320  &  0.957  \\
        \bottomrule
    \end{tabular}}
           \begin{tablenotes}
      \item  {\noindent\footnotesize Note: See the footnote of Table \ref{tab:sim1}.

}
    \end{tablenotes}
\end{table}

 \begin{table}[H]
    \caption{Monte Carlo results for $s = 10$ with $p=100$ covariates ($n=1000$).}
    \label{tab:sim4}
    \centering
    \resizebox{.7\textwidth}{!}{\begin{tabular}{lcccccccc}
        \toprule
        Method & CP & $m$& $|\text{mean bias}|$ & $|\text{median bias}|$ & $\sqrt{\text{Var}}$ & $\text{ E} \sqrt{\text{Var}}$& Cov95 \\
        \midrule
        \multirow{4}{*}{multi-task-debiased-CUE}   & 60  &  15  &  .046  &  .126  &  .902  &  .907  &  0.931  \\
           &120  &  15  &  .054  &  .080  &  .540  &  .422  &  0.927     \\
           &  60  &  30  &  .130  &  .153  &  1.135  &  1.087  &  0.910    \\
           &   120  &  30  &  .100  &  .100  &  .710  &  .690  &  0.922 \\
           \midrule
                \multirow{4}{*}{single-task-debiased-CUE}   & 60  &  15  &  .129  &  .202  &  1.051  &  1.109  &  0.920 \\
            & 120  &  15  &  .132  &  .142  &  .609  &  .553  &  0.915    \\  
            &  60  &  30  &  .207  &  .229  &  1.160  &  1.288  &  0.895 \\
            &    120  &  30  &  .154  &  .186  &  .943  &  1.203  &  0.904 \\ 
        \midrule
                \multirow{4}{*}{TSLS}   &  60  &  15  &  .164  &  .171  &  .192  &  .194  &  0.851   \\
                &120  &  15  &  .107  &  .111  &  .170  &  .163  &  0.875     \\
          &  60  &  30  &  .212  &  .215  &  .156  &  .153  &  0.708    \\
         &    120  &  30  &  .166  &  .166  &  .136  &  .135  &  0.733     \\
                 \midrule
        \multirow{4}{*}{debiased-GMM}   & 60  &  15  &  .208  &  .212  &  .178  &  .185  &  0.781 \\
              & 120  &  15  &  .156  &  .156  &  .158  &  .157  &  0.794    \\
            &   60  &  30  &  .248  &  .250  &  .148  &  .149  &  0.602   \\
            &    120  &  30  &  .211  &  .214  &  .128  &  .132  &  0.618  \\ 
        \midrule
        \multirow{4}{*}{oracle-CUE}   &  60  &  15  &  .046  &  .023  &  .910  &  .988  &  0.959 \\
              &120  &  15  &  .017  &  .003  &  .360  &  .308  &  0.955   \\
          &   60  &  30  &  .040  &  .001  &  1.208  &  1.391  &  0.937 \\ 
          &   120  &  30  &  .038  &  .002  &  .650  &  .690  &  0.950  \\
           \midrule
        \multirow{4}{*}{oracle-GMM}   & 60  &  15  &  .153  &  .155  &  .185  &  .189  &  0.864   \\
            &   120  &  15  &  .098  &  .105  &  .161  &  .156  &  0.867   \\
             & 60  &  30  &  .203  &  .202  &  .151  &  .151  &  0.696   \\
           &  120  &  30  &  .155  &  .160  &  .131  &  .131  &  0.759  \\
           \midrule
        \multirow{4}{*}{Jackknife}    & 60  &  15  &  .581  &  .029  &  18.344  &  234.071  &  0.974 \\
           &     120  &  15  &  .261  &  .010  &  8.706  &  36.470  &  0.956\\
            & 60  &  30  &  .140  &  .051  &  13.819  &  257.802  &  0.974    \\
           &   120  &  30  &  .037  &  .006  &  2.526  &  13.639  &  0.970    \\
        \bottomrule
    \end{tabular}}
           \begin{tablenotes}
      \item  {\noindent\footnotesize Note: See the footnote of Table \ref{tab:sim1}.

}
    \end{tablenotes}
\end{table}

\section{Effect of compulsory school attendance on earnings}
\label{sec:app}

The causal relationship between education and earnings has been a subject of considerable interest. The main concern with causal inference based on observational data is that education level is often not randomly assigned. Hence, potential unmeasured confounders such as individual motivation may affect both education level and earnings. In this section, we revisit the study of \cite{angrist1991does} to estimate the effect of years of education ($D_i$) on log weekly
wage ($Y_i$) in an extract of  $n=329,509$ men born in 1940--1949 across 51 states from the 1980 Census. The study leveraged quarter of birth indicators as IVs, which induced variation in length of education through a combination of school start age policies and compulsory schooling laws, and has been a motivating empirical example in the weak IV literature \citep{bound1995problems, Staiger:1997aa}. In our analysis, the IVs are $m=30$ quarter- and year-of-birth interactions. The covariates are age (measured in quarters of years) and age squared, dummies for marital status, race, center city, eight region-of-residence dummies and nine year-of-birth dummies.
Imposing linearity on the original 22 covariates is a strong assumption. To relax it, we augment the feature set with pairwise interactions among the covariates, yielding 188 variables, and then fit a linear model in this expanded space. Rather than imposing a single pooled relationship across all states, we stratified the sample by state of birth and estimated the model within each stratum.
 
Monte Carlo studies by \cite{hansen2008estimation} suggest that many weak IVs asymptotics may yield improved approximation to finite-sample performance in this application. The 30 instruments are constructed from the same quarter-of-birth and year-of-birth variables, so their dependencies on the covariates are naturally related. This motivates the use of multi-task learning, which allows the nuisance model estimation to borrow strength across instruments components. 
We implement the multi-task debiased-CUE estimator, the single-task counterpart and the conventional TSLS estimator. To allow for heterogeneity across states, we conduct the analysis separately by state of birth, producing a set of state-specific estimates for each method. For readability, Figure 1 reports results for ten representative states. For each state, we plot point estimates and 95\% confidence intervals for the multi-task debiased estimator, the single-task debiased estimator, and conventional TSLS. We draw the following remarks on the results in Figure:
\begin{itemize}
\item In agreement with the Monte Carlo results, debiased-CUE, which accounts for high-dimensional potential confounding factors, has wider 95\% confidence intervals compared to TSLS.
\item The point estimates of debiased-CUE generally agree with those of TSLS across the states. In states like New York, North Carolina, Tennessee, and Rhode Island, the TSLS estimate is significant, while the confidence interval of debiased-CUE includes 0. In states like Connecticut, the estimates from all methods are significant.
\item The confidence intervals of multi-task CUE and single-task CUE are generally similar. However, the point estimates differ. This aligns with our theory and simulation that multi-task CUE is more robust to sparsity assumptions to reduce the bias of the estimate.
\end{itemize}
 
 \begin{figure}[H]
    \centering
    \includegraphics[width=1\linewidth]{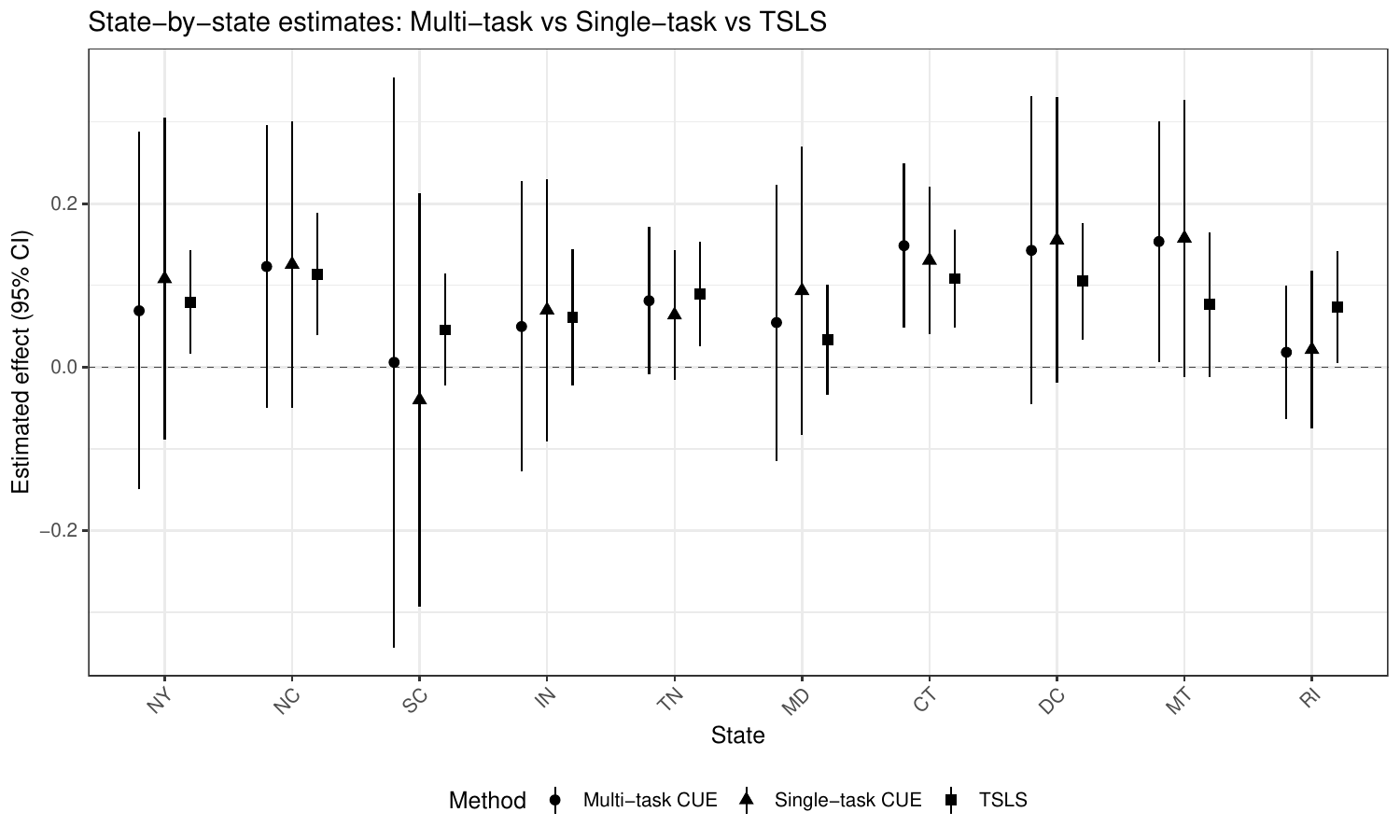}
    \caption{State-specific estimates of the effect of education on earnings for each method}
    \label{fig:state-est}
\end{figure}

\section{Discussion}
\label{sec:discussion}

Many problems in causal inference and missing or coarsened data require first-step estimation of possibly infinite-dimensional nuisance parameters, and  regularized estimation in high-dimensional settings anchored by semiparametric efficiency theory has emerged as one of the leading methods in recent years. In this paper, we extend the framework to the many weak IVs asymptotic regime, in which the number of IVs can diverge with sample size while identification shrinks. We outline some limitations and avenues for future work. While we have considered a univariate treatment, the framework can be extended to incorporate the additive causal effects of multiple treatments. It is also of interest to investigate nonlinear moments which  accommodate binary, count and survival outcomes. In addition, semiparametric efficiency for the many weak IVs asymptotic regime appears to be an open problem, which we hope to investigate in future research.

\section*{Data Availability Statement}

The data that support the findings of this study are openly available in Harvard Dataverse at https://doi.org/doi:10.7910/DVN/ENLGZX.

\section*{Acknowledgements and Disclosure of Funding}

Baoluo Sun's research is partially supported by Singapore MOE AcRF Tier 1 grants (A-8000452-00-00, A-8002935-00-00).



\newpage

\appendix

\section{Proofs}
The organization of the proof is illustrated in Figure \ref{fig:prooffig}.

\begin{figure}[htbp]
    \centering
    \begin{tikzpicture}[>=stealth, node distance=1.5cm]
    \tikzstyle{lemma}=[text=red]
    \node (s1) at (0,4) {Assumption\ref{assum:glmgroupsparse}};
    \node (s2) at (0,3) {Assumption\ref{assum:randomdesign}};
    \node (s3) at (0,2) {Assumption\ref{assum:parameterspace}};
    \node (s4) at (0,1) {Assumption\ref{assum:sparsity}};
    \node (s5) at (0,-1) {Assumption\ref{assum:sparsity2}};
    \node (s6) at (0,0) {Assumption\ref{assum:moments}};

    \node (L1) at (8,4) {Lemma\ref{lemma:conmom}};
    \node (L2) at (8,3) {Lemma\ref{lemma:omega}};
    \node (L3) at (8,2) {Lemma\ref{lemma:moment}};
    \node (L4) at (8,1) {Lemma\ref{lemma:cov}};
    \node (L5) at (8,0) {Lemma\ref{lemma:deQ}};

    \node[align=center] (multi) at (3.5,4) {Theorem \ref{theorem:multirate}};
    \node (c1) at (12,3) {Theorem \ref{theorem::nor}}; 

    \draw[->] (s5) -- (L3);
    \draw[->] (s5) -- (L4); 
    \draw[->] (s5) -- (L5); 

    \draw[->] (s6) -- (L1);
    \draw[->] (s6) -- (L2);
    \draw[->] (s6) -- (L3);
    \draw[->] (s6) -- (L4); 
    \draw[->] (s6) -- (L5); 

    \draw[->] (s1) -- (multi);
    \draw[->] (s2) -- (multi);
    \draw[->] (s3) -- (multi);
    \draw[->] (s4) -- (multi);

    \draw[->] (multi) -- (L3);
    \draw[->] (multi) -- (L4);
    \draw[->] (multi) -- (L5);

    \draw[->] (L1) -- (c1);
    \draw[->] (L2) -- (c1);
    \draw[->] (L3) -- (c1);
    \draw[->] (L4) -- (c1);
    \draw[->] (L5) -- (c1);
    \end{tikzpicture}
    \caption{Organization of the proof}
    \label{fig:prooffig}
\end{figure}

\subsection{Proof of Theorem \ref{theorem:multirate}}
The convergence rate for the usual Lasso estimators $\hat\xi_k,$ and $\hat \gamma_k$ is a standard result in high-dimensional statistics \citep{wainwright2019high} and thus omitted here for brevity.
For $\alpha = (\alpha_1^T,\dots, \alpha_m^T)^T$, denote $||\alpha||_{1,2} = \sum_{l=1}^p (\sum_{j=1}^m \alpha_{jl}^2)^{1/2}$, $||\alpha||_{\infty,2} = \max_{l=1,\dots,p} (\sum_{j=1}^m \alpha_{jl}^2)^{1/2}$, $\mathcal{L}_n (\alpha) := \frac{1}{m(K-1)s}\sum_{j=1}^m \sum_{i\notin I_k}^n\{- Z_{ji} X_i^T\alpha_j + \log (1+e^{X_i^T\alpha_j})\}$. Under this notation, the multi-task GLM estimator objective function (\ref{eq:multitask}) can be written as $\mathcal{L}_n (\alpha) + 2 \lambda_{1,n} ||\alpha||_{1,2} $.
Define the event set $\mathcal{A}(\lambda_{1,n}) = \{ ||\nabla \mathcal{L}_n (\alpha_0)||_{\infty,2} \leq \lambda_{1,n}\}$, and the cone $\mathcal{C}(S,3) :=\{ \Delta: ||\Delta_{S^c}||_{1,2} \leq  3||\Delta_S||_{1,2}\}$. The proof of the theorem is divided into several steps:
\begin{enumerate}
    \item We prove with a high probability that the gradient $\nabla \mathcal{L}_n (\alpha_0)$ lies in the set $\mathcal{A}(\lambda_{1,n})$, when $\lambda_{1,n}$ is in the order of $\frac{1}{\sqrt{mn}}+\frac{\sqrt{\log p}}{m\sqrt{n}}$. 
    \item A restricted eigenvalue condition holds for the GLM model with the group lasso norm.
    \item The estimator error lies in a cone on the event $\mathcal{A}(\lambda_{1,n}).$
    \item Use the cone condition and restricted curvature to derive an $L_2$ error bound for the penalized estimator.
\end{enumerate}
\begin{proof}
    
\noindent \textbf{Step 1:}

\noindent We have $\nabla \mathcal{L}_n (\alpha_0) = \frac{1}{(K-1)s} \sum_{i \notin I_k} L_i(\alpha_0),$ where the random vector $L_i = (L_{i1},\dots,L_{im})^T \in \mathbb{R}^{pm}$ has components,
\begin{align*}
    L_{ijl}(\alpha_0) = \frac{1}{m} \left(\frac{e^{X_i^T \alpha_{0,j}}}{1+e^{X_i^T\alpha_{0,j}}}-Z_{ji} \right)X_{il}, \quad L_{ij} = (l_{ij1},\dots,l_{ijp})^T.
\end{align*}
We let $L_i^{(l)} = (L_{i1l},\dots,L_{iml})^T$ denote the subvector indexed by the dimension index $l$. We then have
\begin{align*}
    ||\nabla \mathcal{L}_n(\alpha_0)||_{\infty,2} := \max_{1\leq l\leq p}\biggr\lVert\frac{1}{(K-1)s}\sum_{i=1}^n 
    L_i^{(l)}\biggr\rVert_2 = \max_{1\leq l\leq p} \sup_{u\in \mathbb{S}^{m-1}} \biggr\langle u, \frac{1}{(K-1)s} \sum_{i=1}^n L_i^{(l)}\biggr\rangle,
\end{align*}
where $\mathbb{S}^{m-1}$ is the Euclidean sphere in $\mathbb{R}^m.$
By Example 5.8 in \cite{wainwright2019high}, we can find a 1/2-covering of $\mathbb{S} = \{u_1,\dots,u_N\}$ in the Euclidean norm with cardinality at most $N\leq 5^m$. For any $u \in \mathbb{S}^{m-1}$, we can find $u_t \in \mathbb{S}$ that $|| u - u_t || \leq 1/2,$ then we have 
\begin{align*}
    \max_{u \in \mathbb{S}^{m-1}} \biggr\langle u, \frac{1}{(K-1)s} \sum_{i=1}^n L_i^{(l)}\biggr\rangle &\leq \max_{t = 1,\dots,N} \biggr\langle u_t, \frac{1}{(K-1)s} \sum_{i=1}^n L_i^{(l)}\biggr\rangle + \biggr\langle u - u_t, \frac{1}{(K-1)s} \sum_{i=1}^n L_i^{(l)}\biggr\rangle \\
    & \leq \max_{t = 1,\dots,N} \biggr\langle u_t, \frac{1}{(K-1)s} \sum_{i=1}^n L_i^{(l)}\biggr\rangle + \frac{1}{2} \biggr\lVert\frac{1}{(K-1)s} \sum_{i=1}^n L_i^{(l)} \biggr\rVert.
\end{align*}
By the variational formulation of the Euclidean norm and rearranging the equation, we have,
\begin{align*}
    ||\nabla \mathcal{L}_n(\alpha_0)||_{\infty,2} \leq 2 \max_{1\leq l\leq p} \max_{t = 1,\dots,N} \biggr\langle u_t, \frac{1}{(K-1)s} \sum_{i=1}^n L_i^{(l)}\biggr\rangle.
\end{align*}
Note that 
\begin{align*}
    \left| X_{il}(\frac{e^{X_i^T \alpha_{0,j}}}{1+e^{X_i^T\alpha_{0,j}}}-Z_{ji})\right| \leq |X_{il}| \max \left\{ \frac{e^{X_i^T \alpha_{0,j}}}{1+e^{X_i^T\alpha_{0,j}}}, 1 -  \frac{e^{X_i^T \alpha_{0,j}}}{1+e^{X_i^T\alpha_{0,j}}}\right\}\leq |X_{il}|.
\end{align*}
By Assumption \ref{assum:randomdesign}, $L_{ijl}$ is zero-mean and sub-Gaussian with parameter at most $\frac{c}{m}$.
 It follows that every coordinate of $\frac{1}{(K-1)s}\sum_{i=1}^n 
    L_i^{(l)}$ is a zero-mean and sub-Gaussian random variable with parameter at most $\frac{c}{m\sqrt{(K-1)s}}.$  
    So the random variable $\langle u_t, \frac{1}{(K-1)s} \sum_{i=1}^n L_i^{(l)}\rangle $ is sub-Gaussian with parameter at most $\frac{c}{m\sqrt{(K-1)s}}.$
Thus, by the sub-Gaussian tail bound and union bound, we have that
\begin{align*}
    P\left(||\frac{1}{(K-1)s}\sum_{i=1}^n L_i^{(l)}||_2 \geq 2t\right) \leq 
    \exp \left(
    -\frac{m^2(K-1)st^2}{2c} + m \log 5
    \right).
\end{align*}
Taking the union over all $p$ groups yields 
\begin{align*}
   P( ||\nabla \mathcal{L}_n(\alpha_0)||_{\infty,2} \geq  2t) \leq 
    \exp \left(
    -\frac{m^2(K-1)st^2}{2c} + m \log 5 + \log p
    \right).
\end{align*}

Then there exist $\lambda_{1,n}$ of order $  \frac{1}{\sqrt{mn}}(1+\frac{\log p}{m})^{1/2}$ that, $P( ||\nabla \mathcal{L}_n(\alpha_0)||_{\infty,2} > \lambda_{1,n}) \leq p^{-c_1}$ for some constants $c_1.$\\

\noindent\textbf{Step 2:}

\noindent We define the sample error in the first-order Taylor expansion of $\mathcal{L}_n$,
\begin{align*}
    \mathcal{E}_n(\Delta) := \mathcal{L}_n (\alpha_0+\Delta) - \mathcal{L}_n (\alpha_0) - \langle\nabla\mathcal{L}_n(\alpha_0), \Delta\rangle.
\end{align*}
By Theorem 9.36 in \cite{wainwright2019high} and Assumption 
4, we have 
\begin{align}
\label{eq:localconvex}
    \mathcal{E}_n(\Delta) \geq \frac{\kappa}{2m} ||\Delta||^2 - c_0 (\sqrt{\frac{1}{(K-1)s}} + \sqrt{\frac{\log pm }{m(K-1)s}} )^2||\Delta||^2_{1,2}, \text{for all } ||\Delta||\leq 1, 
\end{align}
with probability at least $1 - c_1 e^{-c_2(K-1)s},$ where $\kappa$ and $c_0$ are constants independent from $n$, $m$ and $p$. For simplicity of notation, we denote $\tau_n = \sqrt{\frac{1}{(K-1)s}} + \sqrt{\frac{\log pm }{m(K-1)s}} $.\\

\noindent \textbf{Step 3:}

\noindent Denote $\hat{\Delta} = \hat{\alpha} - \alpha_0$, we have $||\alpha_0 + \hat{\Delta}||_{1,2} \geq ||\alpha_{0,S}+\hat{\Delta}_{S^c}||_{1,2}   - ||\hat{\Delta}_{S}||_{1,2} = ||\alpha_{0,S}||_{1,2}+||\hat{\Delta}_{S^c}||_{1,2}   - ||\hat{\Delta}_{S}||_{1,2},$ we then have
\begin{align*}
    ||\alpha_0 + \hat{\Delta}||_{1,2} - ||\alpha_0||_{1,2}  \geq ||\hat{\Delta}_{S^c}||_{1,2}   - ||\hat{\Delta}_{S}||_{1,2}.
\end{align*}
Turning to the cost function difference, we have
\begin{align*}
    \mathcal{L}_n (\hat{\alpha}) - \mathcal{L}_n (\alpha_0) &\geq -|\langle\nabla\mathcal{L}_n (\alpha_0), \hat{\Delta} \rangle| \\
   & \geq - ||\nabla\mathcal{L}_n (\alpha_0)||_{\infty,2} ||\hat{\Delta}||_{1,2} \\
   & \geq - \lambda_{1,n} ||\hat{\Delta}||_{1,2}.
\end{align*}
We thus have 
\begin{align*}
    0 & \geq \mathcal{L}_n (\hat{\alpha}) + 2\lambda_{1,n}||\hat{\alpha}||_{1,2} - \mathcal{L}_n (\alpha_0) - 2\lambda_{1,n}||\alpha_0||_{1,2} \\
    &\geq \lambda_{1,n} (2 ||\hat{\alpha}||_{1,2}  - 2 ||\alpha_0||_{1,2} -  ||\hat{\Delta}||_{1,2})\\
    & \geq \lambda_{1,n}(2 ||\hat{\Delta}_{S^c}||_{1,2} - 2||\hat{\Delta}_S||_{1,2} - ||\hat{\Delta}_{S^c}||_{1,2} - ||\hat{\Delta}_S||_{1,2} ) \\
    & \geq \lambda_{1,n}( ||\hat{\Delta}_{S^c}||_{1,2} - 3||\hat{\Delta}_S||_{1,2} ),
\end{align*}
showing that $\hat{\Delta}$ lies in the cone $\mathcal{C}(S,3)$.\\
 
\noindent\textbf{Step 4:}

\noindent Consider the event set $\mathcal{A}(\lambda_{1,n})$ and parameter $\Delta$ in the set $\mathcal{C}(S,3) \cap \{\Delta\in \mathbb{R}^{mp}: ||\Delta||_2 = R\}$ (value of $R$ would be specified later, but in order for \eqref{eq:localconvex} to hold, we require $R\leq 1$), we establish a lower bound on the function value,
\begin{align*}
    \mathcal{F}(\Delta) := & \mathcal{L}_n (\alpha_0+\Delta) + 2\lambda_{1,n}||\alpha_0 + \Delta||_{1,2} - \mathcal{L}_n (\alpha_0) - 2\lambda_{1,n}||\alpha_0||_{1,2}  \\
    \geq & \langle \nabla \mathcal{L}_n (\alpha_0), \Delta\rangle + \frac{\kappa}{2m} ||\Delta||^2 - c_0\tau_n^2 ||\Delta||_{1,2}^2 + 2 \lambda_{1,n} (||\alpha_0+\Delta||_{1,2} -  ||\alpha_0||_{1,2}) \\
    \geq & \frac{\kappa}{2m} ||\Delta||_2^2 - c_0\tau_n^2 ||\Delta||_{1,2}^2 + 2\lambda_{1,n} ( ||\alpha_0+\Delta||_{1,2} -  ||\alpha_0||_{1,2} - \frac{1}{2}||\Delta||_{1,2}) \\
    \geq & \frac{\kappa}{2m} ||\Delta||_2^2 - c_0\tau_n^2 ||\Delta||_{1,2}^2 +\lambda_{1,n} ( - 3||\Delta_S||_{1,2} ),
\end{align*}
where the first inequality is from Step 2, the second inequality is from the definition of the set $\mathcal{A}(\lambda_{1,n})$, and the third inequality is by the triangle inequality.
Since $||\Delta||_{1,2}^2 \leq (||\Delta_S||_{1,2} + ||\Delta_{S^c}||_{1,2})^2 \leq 16 ||\Delta_S||_{1,2}^2 \leq 16 s_a||\Delta||_2^2$, and $||\Delta_S||_{1,2} \leq \sqrt{s_a}||\Delta||_2$, we obtain,
\begin{align*}
    \mathcal{F}(\Delta) &\geq (\frac{\kappa}{2m} - 16 s_a c_0\tau_n^2)  ||\Delta||_2^2- 3\lambda_{1,n}  \sqrt{s_a} ||\Delta||_2 \\
    & \geq \frac{\kappa}{4m}||\Delta||_2^2 - 3\lambda_{1,n}  \sqrt{s_a} ||\Delta||_2, 
\end{align*}
where the last inequality is by Assumption \ref{assum:sparsity}. 
We then have the function value of $\mathcal{F}$ is greater than 
$0$ when $1  \geq R \geq 12 \frac{m\lambda_{1,n}}{\kappa}\sqrt{s_a}$, by assumption \ref{assum:sparsity}, we then have $12 \frac{m\lambda_{1,n}}{\kappa}\sqrt{s_a} < 1$ for $n$ large enough.
Taking $R = 12 \frac{m\lambda_{1,n}}{\kappa}\sqrt{s_a},$ we have that for all $\Delta \in \mathcal{C}(S,3) \cap \{\Delta\in \mathbb{R}^{mp}: ||\Delta||_2 = R\}$, $\mathcal{F}(\Delta) \geq 0$. If $||\hat{\Delta}|| > R,$ we could find $0 < t < 1$ that $t \hat{\Delta} \in \mathcal{C}(S,3) \cap \{\Delta\in \mathbb{R}^{mp}: ||\Delta||_2 = R\}$.
By the convexity of the objective function $\mathcal{F}$ and the fact that $\mathcal{F}(\hat{\Delta}) \leq 0$, and $\mathcal{F}(0) = 0$, we have $\mathcal{F}(t\hat{\Delta}) \leq t \mathcal{F}(\hat{\Delta}) + (1-t) \mathcal{F}(0)\leq 0 $. This contradicts the fact that $\mathcal{F}(\Delta) \geq 0$ for all $\Delta \in \mathcal{C}(S,3) \cap \{\Delta\in \mathbb{R}^{mp}: ||\Delta||_2 = R\}$. It follows that 
\begin{align*}
    ||\hat{\Delta}|| \leq R  = 12 \frac{m\lambda_{1,n}}{\kappa}\sqrt{s_a} = O(\sqrt{\frac{s_a}{n}} (m+ \log p)^{1/2}).
\end{align*}
\end{proof}

\subsection{Proof of Theorem 2}
We introduce some additional notation in the proof. Let $\textup{tr}(A)$ denote the trace of a matrix $A$, and $\Vert A\Vert_F=\{\textup{tr}(A^\prime A)\}^{1/2}$ its Frobenius norm. For two conformable matrices $A$ and $B$, we write $A\leq B$ if and only if $B-A$ is non-negative definite. Furthermore, we denote 
$$
\begin{aligned}
    &\bar{g}(\beta,\eta)={E}[g_i(\beta,\eta)], \quad
    \bar{Q}(\beta,\eta)=\hat{g}(\beta,\eta)\Omega^{-1}(\beta,\eta_0)\hat{g}(\beta,\eta)/2\\
    &\tilde{g}(\beta)=\frac{1}{n}\sum_{i=1}^n g_i(\beta,{\eta}_0),\quad \tilde{\Omega}(\beta)=\frac{1}{n}\sum_{i=1}^n g_i(\beta,{\eta}_0)g^{\prime}_i(\beta,{\eta}_0).
\end{aligned}$$
To ease presentation, we also consider the linear IV regression model in residualized form,
    \begin{align*}
\bar{Y}_i&=\bar{D}_i\beta_0+\varepsilon_i,\\ \bar{D}_i&=\bar{Z}^{\prime}_{i}\pi+\nu_i,\quad i=1,...,n,
\end{align*}
where  ${E}(\varepsilon_i \mid Z_i , X_i)={E}(\nu_i \mid Z_i , X_i)=0$, $\bar{Z}_i=Z_i-E(Z_i|X_i)$, $\bar{D}_i=D_i-E(D_i|X_i)$, and $\bar{Y}_i=Y_i-E(Y_i|X_i)$. We denote $\Sigma_i$ as the covariance matrix of $(\varepsilon_i,\nu_i)$. The regularity conditions in the following Assumption  \ref{assum:moments} are used to establish consistency and asymptotic normality of the continuous-updating GMM estimator with known $\eta_0$ in \cite{Newey:2009aa}, and imply Lemmas \ref{lemma:conmom} and \ref{lemma:omega} below.  

\begin{assumption}
    \label{assum:moments}
  There is a constant $C$ such that ${E}(\varepsilon^4_i\mid Z_i,X_i)\leq C$, ${E}(\nu^4_i \mid Z_i,X_i)\leq C$, $\lambda_{\min}\left(\Sigma_i\right)\geq 1/C$, $|\bar{Z}^{\prime}_{i}\pi|\leq C$, $\lambda_{\min}({E}(\bar{Z}_i\bar{Z}^{\prime}_i \mid X_i))\geq 1/C,$ and $\lambda_{\max}({E}(\bar{Z}_i \bar{Z}^{\prime}_i\mid X_i))\leq C$  almost surely for each $m\in \mathbb{N}^{+}$.  Furthermore, $m^3/n\rightarrow 0$.
\end{assumption}

\setcounter{theorem}{0}

\begin{lemma}
\phantomsection\label{lemma:conmom}
    Under Assumption \ref{assum:moments}, there is a constant $C$ such that 
    \begin{enumerate}
        \item[(i)] ${E}\left(\bar{D}_i^2\mid X_i\right)\leq C$;
        \item[(ii)] $C^{-1}\leq{E}\{\left(\bar{Y}_i-\bar{D}_i\beta\right)^2\bigr\rvert  Z_i, X_i\}\leq C$ almost surely, for all $\beta\in \mathcal B$.
    \end{enumerate}
\end{lemma}
\begin{lemma}
\phantomsection\label{lemma:omega}
Under Assumption \ref{assum:moments}, there is a constant $C$ such that 
\begin{enumerate}
    \item[(i)] $1/C\leq\lambda_{\min}(\Omega(\beta,\eta_0)),\lambda_{\max}(\Omega(\beta,\eta_0))\leq C$ for all $\beta\in\mathcal B$;
    \item[(ii)] $\lambda_{\max}({E}(G_iG^\prime_i))\leq C$;
    \item[(iii)] $||{E}(g_i(\beta,\eta_0)G_i^\prime)||\leq C$, for all $\beta\in\mathcal B$.
\end{enumerate}
\end{lemma}

\noindent Next, we prove some key equicontinuity
conditions for consistency and asymptotic normality of the continuous-updating GMM estimator with first-stage regularized estimation. 

\begin{lemma}
\phantomsection\label{lemma:moment}
    Under Assumptions 1--6, the following equicontinuity conditions hold:
    \begin{enumerate}
        \item[(i)] $\Vert \hat{g}(\beta_0,\hat{\eta})-\tilde{g}(\beta_0)\Vert=o_p(n^{-1/2})$;
        \item[(ii)] $\Vert \hat{G}(\hat{\eta})-\hat{G}(\eta_0)\Vert=o_p(n^{-1/2})$;
        \item[(iii)] $\underset{\beta\in \mathcal B}{\sup}\Vert \hat{g}(\beta,\hat{\eta})-\tilde{g}(\beta)\Vert=o_p(n^{-1/2})$.
    \end{enumerate}
\end{lemma}
\begin{proof}
By the triangle inequality,
\begin{align*}
    &\sqrt{n}\Vert \hat{g}(\beta_0,\hat{\eta})-\tilde{g}(\beta_0)\Vert\\
    = &\sqrt{n}\Vert \frac{1}{n} \underset{k\in[K]}{\sum} \sum_{i\in I_k} (Z_i - \hat{\alpha}_k(X_i) )(Y_i - \hat{\ell}_k (X_i) - (D_i - \hat{r}_k (X_i) \beta_0 ) - \bar{Z}_i(\bar{Y}_i - \bar{D}_i \beta_0)\Vert \\
    \leq&\underset{k\in[K]}{\sum}\underbrace{\sqrt{n}\biggr\lVert\frac{1}{n}\sum_{i\in I_k}\left\{\hat{\alpha}_{k}(X_i)-{\alpha}_{0}(X_i)\right\}\left\{\hat{\ell}_k(X_i)-\ell_0(X_i)-(\hat{r}_k(X_i)-r_0(X_i))\beta_0\right\}\biggr\rVert}_{:=a_k}\\
&+\underset{k\in[K]}{\sum}\underbrace{\sqrt{n}\biggr\lVert\frac{1}{n}\sum_{i\in I_k}\bar{Z}_i\left\{\hat{\ell}_k(X_i)-\ell_0(X_i)-(\hat{r}_k(X_i)-r_0(X_i))\beta_0\right\}\biggr\rVert}_{:=b_k}\\
&+\underset{k\in[K]}{\sum}\underbrace{\sqrt{n}\biggr\lVert\frac{1}{n}\sum_{i\in I_k}\left\{\hat{\alpha}_{k}(X_i)-{\alpha}_{0}(X_i)\right\}(\bar{Y}_i-\bar{D}_i\beta_0)\biggr\rVert}_{:=c_k}.
\end{align*}
For term $a_k$, by Cauchy-Schwarz inequality for conditional expectation  we have
\begin{align*}
    &{E}\left\{a_k\bigr\rvert (O_i)_{i\in I_k^c}\right\}\\
    \leq&{E}\biggr\{\frac{\sqrt{n}}{n}\underset{i\in I_k}{\sum}\Vert(\hat{\alpha}_k(X_i)-\alpha_0(X_i))[\hat{\ell}_k(X_i)-\ell_0(X_i)-(\hat{r}_k(X_i)-r_0(X_i))\beta_0]\Vert\mid (O_i)_{i\in I_k^c}\biggr\}\\
    =&\sqrt{n}{E}\biggr\{\Vert (\hat{\alpha}_k(X_i)-\alpha_0(X_i))[\hat{\ell}_k(X_i)-\ell_0(X_i)-(\hat{r}_k(X_i)-r_0(X_i))\beta_0]\Vert\mid (O_i)_{i\in I_k^c}\biggr\}\\
    \leq&\sqrt{n}{E}^{1/2}\biggr\{ \Vert \hat{\alpha}_k(X_i)-\alpha_0(X_i)\Vert^2\mid (O_i)_{i\in I_k^c}\biggr\}{E}^{1/2}\biggr\{[\hat{\ell}_k(X_i)-\ell_0(X_i)-(\hat{r}_k(X_i)-r_0(X_i))\beta_0]^2\mid (O_i)_{i\in I_k^c}\biggr\}\\
    =&o_p(1),
\end{align*}
where the last equality holds by Assumptions \ref{assum:glmgroupsparse}--\ref{assum:sparsity2}. Then by Lemma 6.1 in \cite{Chernozhukov2018ddml}, the conditional convergence in probability to zero implies $a_k$ is also $o_p(1)$. Next, by Jensen's inequality we have,
\begin{align*}
    &{E}\left\{b_k\bigr\rvert (O_i)_{i\in I^c_k}\right\}\\
    \leq&{E}^{1/2}\biggr\{ \frac{n}{n^2}\Vert \underset{i\in I_k}{\sum}\bar{Z}_i[\hat{\ell}_k(X_i)-\ell_0(X_i)-(\hat{r}_k(X_i)-r_0(X_i))\beta_0]\Vert^2\mid (O_i)_{i\in I^c_k}\biggr\}\\
    =&\frac{s}{n}{E}^{1/2}\{\Vert \bar{Z}_i[\hat{\ell}_k(X_i)-\ell_0(X_i)-(\hat{r}_k(X_i)-r_0(X_i))\beta_0]\Vert^2\mid (O_i)_{i\in I^c_k}\}\\
    =&\frac{s}{n}{E}^{1/2}\{ {E}[\Vert \bar{Z}_i\Vert^2\mid X_i][\hat{\ell}_k(X_i)-\ell_0(X_i)-(\hat{r}_k(X_i)-r_0(X_i))\beta_0]^2\mid (O_i)_{i\in I_k^c}\}\\
    \leq& C\sqrt{m}{E}^{1/2}\{ [\hat{\ell}_k(X_i)-\ell_0(X_i)-(\hat{r}_k(X_i)-r_0(X_i))\beta_0]^2\mid (O_i)_{i\in I_k^c}\}\\
    =&o_p(1),
\end{align*}
where the second equality holds because the nuisance estimators could be treated as being fixed conditioning on the data $(O_i)_{I_{k}^c}$ and ${E}\left(\Vert \bar{Z}_i\Vert^2\mid X_i\text{, }(O_i)_{i\in I_k^c}\right)={E}\left(\Vert \bar{Z}_i\Vert^2\mid X_i\right)$ for each $i\in I_k$ by identical and independently distributed data, and the last equality holds by Assumptions \ref{assum:glmgroupsparse}--\ref{assum:sparsity2}. For term $c$, similar to the analysis for $b_k$, we have
${E}\left\{ c_k\bigr\rvert(O_i)_{i\in I_k^c}\right\} = o_p(1).$ By Lemma 6.1 in \cite{Chernozhukov2018ddml}, terms $a_k$, $b_k$ and $c_k$ are all $o_p(1)$, and the proof of (i) follows by $K$ is a finite number. Although details are omitted, the proof of (ii) is similar to that of (i). Finally, (iii) follows from (i), (ii), the linearity of $\hat{g}(\beta,\eta)$ with respect to $\beta$ and the compactness of $\mathcal B$.

\end{proof}
\setcounter{assumption}{3}

\begin{lemma}
    \label{lemma:cov}
    Under Assumptions 1--6, the following equicontinuity conditions hold:
    \begin{enumerate}
        \item[(i)] $m^{1/2}\Vert \hat{\Omega}(\beta_0,\hat{\eta})-\Omega(\beta_0,\eta_0)\Vert=o_p(1)$;
        \item[(ii)] $m^{1/2}\bigr\Vert n^{-1}\underset{i=1}{\overset{n}{\sum}}g_i(\beta_0,\hat{\eta})G_i(\hat{\eta})^\prime-{E}(g_iG_{i}^\prime)\bigr\Vert=o_p(1)$;
        \item[(iii)] $m^{1/2}\bigr\Vert n^{-1}\underset{i=1}{\overset{n}{\sum}}G_{i}(\hat{\eta})G_{i}(\hat{\eta})^\prime-{E}(G_{i}G_{i}^\prime)\bigr\Vert=o_p(1)$;
        \item[(iv)] $m^{1/2}\underset{\beta\in \mathcal B}{\sup}\Vert \hat{\Omega}(\beta,\hat{\eta})-\Omega(\beta,\eta_0)\Vert=o_p(1)$;
        \item[(v)] $m^{1/2}\underset{\beta\in \mathcal B}{\sup}\bigr\Vert n^{-1}\underset{i=1}{\overset{n}{\sum}}g_i(\beta,\hat{\eta})G_{i}(\hat{\eta})^\prime-{E}\{g_i(\beta,\eta_0)G_{i}^\prime\}\bigr\Vert=o_p(1)$.
    \end{enumerate}
\end{lemma}
\begin{proof}
   (i) By triangle inequality,
    \begin{align*}
        &\Vert \hat{\Omega}(\beta_0,\hat{\eta})-\Omega(\beta_0,\eta_0)\Vert
        \leq\Vert \hat{\Omega}(\beta_0,\hat{\eta})-\tilde{\Omega}(\beta_0)\Vert+\Vert \tilde{\Omega}(\beta_0)-\Omega(\beta_0,\eta_0)\Vert.
    \end{align*}
By direct expansion of the first term on the right hand side and triangle inequality, we have
\begin{align*}
    &\Vert \hat{\Omega}(\beta_0,\hat{\eta})-\tilde{\Omega}(\beta_0)\Vert\\
    \leq&\biggr\Vert \frac{1}{n}\underset{i\in [n]}{\sum}\bar{Z}_i(\hat{\alpha}(X_i)-\alpha_0(X_i))^\prime(\bar{Y}_i-\bar{D}_i\beta_0)^2\biggr\Vert+\biggr\Vert \frac{1}{n}\underset{i\in [n]}{\sum}(\hat{\alpha}(X_i)-\alpha_0(X_i))\bar{Z}_i^\prime(\bar{Y}_i-\bar{D}_i\beta_0)^2\biggr\Vert\\
    &+2\biggr\Vert \frac{1}{n}\underset{i\in [n]}{\sum}\bar{Z}_i\bar{Z}_i^\prime(\bar{Y}_i-\bar{D}_i\beta_0)(\hat{\ell}(X_i)-\ell_0(X_i))\biggr\Vert+2\beta_0\biggr\Vert \frac{1}{n}\underset{i\in [n]}{\sum} \bar{Z}_i\bar{Z}_i^\prime(\bar{Y}_i-\bar{D}_i\beta_0)(\hat{r}(X_i)-r_0(X_i))\biggr\Vert\\
    &+\textup{Rem}_{\Omega}\\
    =&\Vert I_1\Vert+\Vert I_2\Vert+ 2 \Vert I_3\Vert+\Vert I_4\Vert+\textup{Rem}_{\Omega},
\end{align*}
where the $\textup{Rem}_{\Omega}$ comprises of higher order estimation error terms, so we focus on the first four terms. For $ i \in I_k$, denote $H_i = \bar{Z}_i(\hat{\alpha}_k(X_i)-\alpha_0(X_i))^\prime(\bar{Y}_i-\bar{D}_i\beta_0)^2$. For $\Vert I_1\Vert$, by triangle inequality, we have
\begin{align}
\label{eq:decomposition}
     \Vert I_1\Vert  
    \leq&\underset{k\in[K]}{\sum}\biggr(\biggr\Vert \frac{1}{n}\underset{i\in I_k}{\sum}\{ H_i -{E}[H_i\mid \{O_i\}_{i\in I_k^c}]\}\biggr\Vert+\biggr\Vert \frac{s}{n}{E}[H_i\mid \{O_i\}_{i\in I_k^c}]\biggr\Vert\biggr), 
\end{align}
where $H_i-{E}[H_i\mid (O_i)_{i\in I_k^c}]$ are independent with mean zero conditioning on observed data $(O_i)_{i\in I_k^c}$. Because $K$ is a finite number, it is enough to consider one specific $k\in\{1,...,K\}$. By the conditional independency as in the proof of Lemma \ref{lemma:moment}, we have
\begin{align*}
    &{E}\biggr\{\biggr\Vert\frac{1}{n}\underset{i\in I_k}{\sum} H_i-{E}[H_i\mid (O_i)_{i\in I_k^c}]\biggr\Vert^2\mid (O_i)_{i\in I_k^c}\biggr\}\\
    \leq&{E}\biggr\{\biggr\Vert\frac{1}{n}\underset{i\in I_k}{\sum} H_i-{E}[H_i\mid (O_i)_{i\in I_k^c}]\biggr\Vert^2_F\mid(O_i)_{i\in I_k^c}\biggr\}\\
    =&\frac{s}{n^2}{E}\{\Vert H_i-{E}[H_i|(O_i)_{i\in I_k^c}]\Vert^2_F\mid(O_i)_{i\in I_k^c}\}\\
    =& o_p(n^{-1}),
\end{align*}
where the first inequality is from the fact that the spectral norm of a matrix is less than or equal  to its Frobenius norm. The second  equality holds because 
\begin{align*}
    &{E}\{\Vert H_i-{E}[H_i\mid(O_i)_{i\in I_k^c}]\Vert_F^2\mid (O_i)_{i\in I_k^c}\}\\
    \leq& {E}\{\Vert H_i\Vert_F^2|(O_i)_{i\in I_k^c}\}\\
    =&{E}\{\Vert \bar{Z}_i(\hat{\alpha}_k(X_i)-\alpha_0(X_i))^\prime(\bar{Y}_i-\bar{D}_i\beta_0)^2\Vert_F^2\mid(O_i)_{i\in I_k^c}\}\\
    =&{E}\{\Vert \bar{Z}_i(\hat{\alpha}_k(X_i)-\alpha_0(X_i))^\prime\Vert_F^2{E}[(\bar{Y_i}-\bar{D}_i\beta_0)^4\mid X_i,Z_i]\mid (O_i)_{i\in I_k^c}\}\\
    \leq&C{E}\{ \Vert \bar{Z}_i\Vert^2\Vert\hat{\alpha}_k(X_i)-\alpha_0(X_i)\Vert^2\mid (O_i)_{i\in I_k^c}\}\\
        =&C{E}\{ {E}[\Vert \bar{Z}_i\Vert^2\mid X_i]\Vert\hat{\alpha}_k(X_i)-\alpha_0(X_i)\Vert^2\mid (O_i)_{i\in I_k^c}\}\}\\
    =&o_p(1),
\end{align*}
where the second inequality is from Assumption \ref{assum:moments}, and the last equality is from Assumptions \ref{assum:glmgroupsparse}--\ref{assum:sparsity2}. For the second term on the right-hand side of (\ref{eq:decomposition}), we have
\begin{align}
    &\bigr\Vert \frac{s}{n}{E}\{\bar{Z}_i(\hat{\alpha}_k(X_i)-\alpha_0(X_i))^\prime(\bar{Y}_i-\bar{D}_i\beta_0)^2\mid \{O_i\}_{i\in I_k^c}\}\bigr\Vert^2\nonumber\\
    \leq& C \underset{j\in [m]}{\sum}\Vert {E}\{ \bar{Z}_i(\hat{\alpha}_{j,k}(X_i)-\alpha_{j,0}(X_i))(\bar{Y}_i-\bar{D}_i\beta_0)^2\mid \{O_i\}_{i\in I_k^c}\}\Vert^2    \label{equ::omega}.
\end{align}
By the conditional version of matrix Cauchy-Schwarz inequality in \cite{tripathi1999matrix}, we have
\begin{align*}
    &{E}\{\bar{Z}_i(\hat{\alpha}_{j,k}(X_i)-\alpha_{j,0}(X_i))(\bar{Y}_i-\bar{D}_i\beta_0)^2\mid \{O_i\}_{i\in I_k^c}\}{E}\{\bar{Z}_i^\prime(\hat{\alpha}_{j,k}(X_i)-\alpha_{j,0}(X_i))(\bar{Y}_i-\bar{D}_i\beta_0)^2\mid \{O_i\}_{i\in I_k^c}\}\\
    \leq& {E}[(\hat{\alpha}_{j,k}(X_i)-\alpha_{j,0}(X_i))^2(\bar{Y}_i-\bar{D}_i\beta_0)^2\mid \{O_i\}_{i\in I_k^c}]{E}[\bar{Z}_i\bar{Z}_i^\prime(\bar{Y}_i-\bar{D}_i\beta_0)^2]\quad\\
    =& {E}[(\hat{\alpha}_{j,k}(X_i)-\alpha_{j,0}(X_i))^2{E}[(\bar{Y}_i-\bar{D}_i\beta_0)^2\mid Z_i,X_i]\mid \{O_i\}_{i\in I_k^c}]{E}[\bar{Z}_i\bar{Z}_i^\prime(\bar{Y}_i-\bar{D}_i\beta_0)^2]\\
    \leq&C{E}[(\hat{\alpha}_{j,k}(X_i)-\alpha_{j,0}(X_i))^2\mid\{O_i\}_{i\in I_k^c}]\Omega(\beta_0,\eta_0),
\end{align*}
where the last inequality is from Assumption \ref{assum:moments}. Applying this result to (\ref{equ::omega}), we have
\begin{align*}
    &\underset{j\in [m]}{\sum}\Vert {E}\{ \bar{Z}_i(\hat{\alpha}_{j,k}(X_i)-\alpha_{j,0}(X_i))(\bar{Y}_i-\bar{D}_i\beta_0)^2\mid \{O_i\}_{i\in I_k^c}\}\Vert^2\\
    \leq&C\underset{j\in [m]}{\sum}{E}[( \hat{\alpha}_{j,k}(X_i)-\alpha_{j,0}(X_i))^2\mid\{O_i\}_{i\in I_k^c}]\Vert \Omega(\beta_0,\eta_0)\Vert\\
    \leq&C{E}[\Vert \hat\alpha_k(X_i)-\alpha_0(X_i)\Vert^2\mid \{O_i\}_{i\in I_k^c}] \\
    =&o_p(m^{-1}),
\end{align*}
where the last equality is from Assumptions \ref{assum:glmgroupsparse}--\ref{assum:sparsity2}. Then by Lemma 6.1 in \cite{Chernozhukov2018ddml}, we have ${m}^{1/2}\Vert I_1\Vert=o_p(1)$. Because the spectral norm remains the same after taking transpose, $\Vert I_2\Vert=\Vert I_1\Vert$. Next, we consider $I_3$. Since the conditional zero mean condition ${E}\{\bar{Z}_i\bar{Z}_i^\prime(\bar{Y}_i-\bar{D}_i\beta_0)(\hat{\ell}(X_i)-\ell_0(X_i))|(O_i)_{i\in I_k^c}\}=0$ holds, similar to the analysis for $\Vert I_1\Vert$, we have
\begin{align*}
    &{E}\{\Vert I_3\Vert^2\mid (O_i)_{i\in I_k^c}\}\\
    \leq&\underset{k\in[K]}{\sum}\frac{s}{n^2}{E}\{ \Vert \bar{Z}_i\bar{Z}_i^\prime\Vert^2(\hat{\ell}(X_i)-\ell_0(X_i))^2{E}[(\bar{Y}_i-\bar{D}_i\beta_0)^2\mid Z_i,X_i]\mid (O_i)_{i\in I_k^c}\}\\
    \leq& \underset{k\in[K]}{\sum}\frac{Cs}{n^2}{E}\{{E}[\Vert \bar{Z}_i\Vert^4\mid X_i](\hat{\ell}(X_i)-\ell_0(X_i))^2\mid (O_i)_{i\in I_k^c}\}\\
    =&o_p(mn^{-1}),
\end{align*}
where the first inequality is from triangle inequality, the second inequality is from Assumption \ref{assum:moments} and the proof of Lemma \ref{lemma:moment}, and the last equality is from Theorem \ref{theorem:multirate}, Assumption \ref{assum:parameterspace}, \ref{assum:sparsity2}. Then by Lemma 6.1 in \cite{Chernozhukov2018ddml}, we have $\sqrt{m}\Vert I_3\Vert=o_p(1)$. The term $\Vert I_4\Vert$ can be bounded in a similar way as $\Vert I_3\Vert$, so we can conclude that $m^{1/2}\Vert \hat{\Omega}(\beta_0,\hat{\eta})-\tilde{\Omega}(\beta_0)\Vert=o_p(1)$. Moreover, $\text{Rem}_{\Omega}$ is negligible relative to the previous terms, it will be also $o_p(1)$. The remaining term satisfies $m^{1/2}\Vert \tilde{\Omega}(\beta_0)-\Omega(\beta_0,\eta_0)\Vert=o_p(1)$ by Lemma A11 in \cite{Newey:2009aa}. Lemma \ref{lemma:cov} (ii) and (iii) could be proved in a similar way as (i) and their proofs are thus omitted.\\

\noindent (iv) By using the fact that $g_i(\beta,\eta)=g_i(\beta_0,\eta)+G_i(\eta)(\beta-\beta_0)$,
\begin{align*}
    &\underset{\beta\in \mathcal B}{\sup}\Vert \hat{\Omega}(\beta,\hat{\eta})-\Omega(\beta,\eta_0)\Vert\\
    =& \underset{\beta\in \mathcal B}{\sup}\biggr\Vert n^{-1}\underset{i\in [n]}{\sum}(g_i(\beta_0,\hat{\eta})+G_i(\hat{\eta})(\beta-\beta_0))(g_i(\beta_0,\hat{\eta})+G_i(\hat{\eta})(\beta-\beta_0))^\prime\\
    &-{E}[(g_i(\beta_0,\eta_0)+G_i(\eta_0)(\beta-\beta_0))(g_i(\beta_0,\eta_0)+G_i(\eta_0)(\beta-\beta_0))^\prime]\biggr\Vert\\
    \leq&\Vert \hat{\Omega}(\beta_0,\hat{\eta})-\Omega\Vert+ 2\underset{\beta\in \mathcal B}{\sup}|\beta-\beta_0|\Vert n^{-1}\underset{i\in[n]}{\sum}g_i(\beta_0,\hat{\eta})G_i(\hat{\eta})^\prime-{E}[g_i(\beta_0,\eta_0)G_i(\eta_0)^\prime]\Vert\\
    &+\underset{\beta\in \mathcal B}{\sup}(\beta-\beta_0)^2\biggr\Vert n^{-1}\underset{i\in[n]}{\sum} G_i(\hat{\eta})G_i(\hat{\eta})^\prime-{E}[G_i(\eta_0)G_i(\eta_0)^\prime]\biggr\Vert.
\end{align*}
The result then holds by Lemma \ref{lemma:cov} (i)--(iii) and the compactness of $\mathcal B$. We omit the proof of (v) as it is similar to that of (iv).
\end{proof}
\begin{lemma}
    \label{lemma:deQ}
    Under Assumptions 1-6, 
    \begin{align*}
       &\text{(i) }  n\mu_n^{-1}\biggr|\frac{\partial\hat{Q}(\beta,\hat{\eta})}{\partial\beta}\biggr\rvert_{\beta=\beta_0}-\frac{\partial \bar{Q}(\beta,\hat{\eta})}{\partial\beta}\biggr\rvert_{\beta=\beta_0}\biggr|=o_p(1)\\
       &\text{(ii) } n\mu_n^{-2}\underset{\beta\in \mathcal{B}}{\sup} \biggr|\frac{\partial^2\hat{Q}(\beta,\hat{\eta})}{\partial \beta^2}-\frac{\partial^2\bar{Q}(\beta,\hat{\eta})}{\partial \beta^2}\biggr|=o_p(1)
    \end{align*}
\end{lemma}
Lemma \ref{lemma:deQ} follows from the key equicontinuity conditions established in Lemmas \ref{lemma:conmom}--\ref{lemma:cov}. We omit the proof of Lemma \ref{lemma:deQ} and the remaining proof of Theorem \ref{theorem::nor} for brevity, which follows closely our earlier preprint \citep{zhang2025debiasedcontinuousupdatinggmm} and \citet{ye2024genius}.

\subsection{Proof of Theorem 3}
\begin{proof}
    We first show that the variance estimator is consistent, $\mu_n^2\hat{V}/n\overset{p}{\rightarrow}V$. Let
    $$ \hat{D}(\beta,\eta)=\frac{\partial \hat{g}(\beta,\eta)}{\partial \beta}-\frac{1}{n}\sum_{i=1}^n\{G_i(\eta)g_i(\beta,\eta)^\prime\}\hat{\Omega}^{-1}(\beta,\eta)\hat{g}(\beta,\eta).$$
    We denote 
    \begin{align*}
      \hat{V}(\beta,\eta)=\left\{\frac{\partial^2 \hat{Q}(\beta,\eta)}{\partial \beta^2}\right\}^{-2}\hat{D}(\beta,\eta)^\prime\hat{\Omega}^{-1}(\beta,\eta)\hat{D}(\beta,\eta),
    \end{align*} 
and $\bar{\beta}=\underset{\beta\in \mathcal B}{\argmin}\bar{Q}(\beta,\eta_0)$. Our goal is to show $\mu_n^2\hat{V}/n-\mu_n^2\hat{V}(\bar{\beta},\eta_0)/n=o_p(1)$. Firstly, by decomposition we have
\begin{align*}
    &\mu_n^2\hat{V}/n-\mu_n^2\hat{V}(\bar{\beta},\eta_0)/n\\
    =&[(n\mu_n^{-2}\frac{\partial^2\hat{Q}(\hat{\beta},\hat{\eta})}{\partial\beta^2})^{-2}-(n\mu_n^{-2}\frac{\partial^2\hat{Q}(\bar{\beta},\eta_0)}{\partial\beta^2})^{-2}]\mu_n^{-1}n^{1/2}\hat{D}^\prime\hat{\Omega}^{-1}(\hat{\beta},\hat{\eta})\mu_n^{-1}n^{1/2}\hat{D}\\
    +&(n\mu_n^{-2}\frac{\partial^2\hat{Q}(\bar{\beta},\eta_0)}{\partial\beta^2})^{-2}\mu_n^{-1}n^{1/2}[\hat{D}-\hat{D}(\bar{\beta},\eta_0)]^\prime\hat{\Omega}^{-1}(\hat{\beta},\hat{\eta})\mu_n^{-1}n^{1/2}\hat{D}\\
    +&(n\mu_n^{-2}\frac{\partial^2\hat{Q}(\bar{\beta},\eta_0)}{\partial\beta^2})^{-2}\mu_n^{-1}n^{1/2}\hat{D}(\bar{\beta},\eta_0)^\prime(\hat{\Omega}^{-1}(\hat{\beta},\hat{\eta})-\hat{\Omega}^{-1}(\bar{\beta},\eta_0))\mu_n^{-1}n^{1/2}\hat{D}\\
    +&(n\mu_n^{-2}\frac{\partial^2\hat{Q}(\bar{\beta},\eta_0)}{\partial\beta^2})^{-2}\mu_n^{-1}n^{1/2}\hat{D}(\bar{\beta},\eta_0)^\prime\hat{\Omega}^{-1}(\bar{\beta},\eta_0)\mu_n^{-1}n^{1/2}[\hat{D}-\hat{D}(\bar{\beta},\eta_0)]\\
    =&A_1+A_2+A_3+A_4.
\end{align*}
We show that each of the terms $A_1$, $A_2$, $A_3$ and $A_4$ is $o_p(1)$. First, by the triangle inequality,
\begin{align*}
    &\Vert \hat{\Omega}^{-1}(\hat{\beta},\hat{\eta})\Vert\\
    \leq&\Vert \hat{\Omega}^{-1}(\beta_0,\hat{\eta})\Vert+\Vert \hat{\Omega}^{-1}(\beta_0,\hat{\eta})-\hat{\Omega}^{-1}(\hat{\beta},\hat{\eta})\Vert\\
    =& \Vert \hat{\Omega}^{-1}(\beta_0,\hat{\eta})\Vert+\Vert\hat{\Omega}^{-1}(\hat{\beta},\hat{\eta})\{\hat{\Omega}(\hat\beta,\hat\eta)-\hat\Omega(\beta_0,\hat\eta)\}\hat\Omega^{-1}(\beta_0,\hat\eta)\Vert\\
    \leq& O_p(1)+C|\beta_0-\hat{\beta}|\biggr\Vert n^{-1}\underset{i=1}{\overset{n}{\sum}}g_i(\beta_0,\hat{\eta})G_i(\hat\eta)^\prime\biggr\Vert+ C|\beta_0-\hat{\beta}|^2\biggr\Vert n^{-1}\underset{i=1}{\overset{n}{\sum}}G_i(\hat{\eta})G^\prime_i(\hat{\eta})\biggr\Vert\quad w.p.1\\
    =&O_p(1),
\end{align*}
where the last inequality  holds because $\hat{\beta}\overset{p}{\rightarrow}\beta$ and Lemma \ref{lemma:omega}. For the same reason, we could have $\Vert \hat{\Omega}^{-1}(\bar{\beta},\eta_0)\Vert\leq C$. Next, by the triangle inequality we could verify that,
 \begin{align*}
     &\Vert \frac{\sqrt{n}}{\mu_n}[\hat{D}-\hat{D}(\bar{\beta},\eta_0)]\Vert\\
     \leq& \Vert \frac{\sqrt{n}}{\mu_n}\{\hat{D}(\hat{\beta},\hat\eta)- \hat{D}(\hat{\beta},\eta_0)\}\Vert+\Vert \frac{\sqrt{n}}{\mu_n}\{\hat{D}(\hat{\beta},\eta_0)-\hat{D}(\bar{\beta},\eta_0)\}\Vert\\
     \leq&\Vert \frac{\sqrt{n}}{\mu_n}\{\hat{D}(\hat{\beta},\hat\eta)- \hat{D}(\hat{\beta},\eta_0)\}\Vert\\
     &+\frac{\sqrt{n}}{\mu_n}|\hat{\beta}-\bar{\beta}| \biggr\Vert n^{-1}\underset{i=1}{\overset{n}{\sum}}\{G_i(\eta_0)G^\prime_i(\eta_0)\}\hat{\Omega}^{-1}(\hat{\beta},\eta_0)\hat{g}(\hat{\beta},\eta_0)\biggr\Vert\\
     &+\frac{\sqrt{n}}{\mu_n}\biggr\Vert n^{-1}\underset{i=1}{\overset{n}{\sum}}\{G_i(\eta_0)g^\prime_i(\bar{\beta},\eta_0)\}[\hat\Omega(\hat\beta,\eta_0)-\hat\Omega(\bar{\beta},\eta_0)]\hat{g}(\hat{\beta},\eta_0)\biggr\Vert\\
     &+\frac{\sqrt{n}}{\mu_n}|\hat{\beta}-\bar{\beta}|\biggr\Vert n^{-1}\underset{i=1}{\overset{n}{\sum}}\{G_i(\eta_0)g^\prime_i(\bar{\beta},\eta_0)\hat{\Omega}^{-1}(\bar{\beta},\eta_0)\}\hat{G}(\eta_0)\biggr\Vert\\
     \leq&\Vert \frac{\sqrt{n}}{\mu_n}\{\hat{D}(\hat{\beta},\hat\eta)- \hat{D}(\hat{\beta},\eta_0)\}\Vert+o_p(1)\\
     \leq&\Vert \frac{\sqrt{n}}{\mu_n}\{\hat{D}(\hat{\beta},\hat\eta)- \hat{D}(\beta_0,\hat{\eta})\}\Vert+\Vert \frac{\sqrt{n}}{\mu_n}\{\hat{D}(\hat{\beta},\eta_0)- \hat{D}(\beta_0,\eta_0)\}\Vert\\
     &+\Vert \frac{\sqrt{n}}{\mu_n}\{ \hat{D}(\beta_0,\hat{\eta})-\hat{D}(\beta_0,\eta_0)\}\Vert+o_p(1)\\
     =&o_p(1),
 \end{align*}
where both $\hat{\beta}$ and $\bar{\beta}$ are the consistent estimators of $\beta_0$, and we make use of the fact that 
\begin{align}
\label{eq:hatgdecompose}
    \tilde{g}(\beta)&=\tilde{g}(\beta_0)+(\beta-\beta_0)\hat{G}(\eta_0),\\
\label{eq:gidecompose}    
    g_i(\beta,\eta_0)&=g_i(\beta_0,\eta_0)+(\beta-\beta_0)G_i(\eta_0),
\end{align}
which yields the third inequality by Lemma \ref{lemma:omega}. 
Lastly, by the triangle inequality again we have
\begin{align*}
    &\Vert \frac{\sqrt{n}}{\mu_n}\hat D(\hat{\beta},\hat{\eta})\Vert\\
    \leq& \frac{\sqrt{n}}{\mu_n}\Vert \hat{G}(\hat{\eta})\Vert+\frac{\sqrt{n}}{\mu_n}\biggr\Vert n^{-1}\underset{i=1}{\overset{n}{\sum}}\{G_i(\hat{\eta})g^\prime_i(\hat{\beta},\hat{\eta})\}\hat{\Omega}^{-1}(\hat{\beta},\hat{\eta})\hat{g}(\hat{\beta},\hat{\eta})\biggr\Vert\\
    =&O_p(1),
\end{align*}
where the equality holds by (\ref{eq:hatgdecompose}), (\ref{eq:gidecompose}), and Lemmas \ref{lemma:omega}--\ref{lemma:cov}. Following similar steps, we could also show $\Vert \frac{\sqrt{n}}{\mu_n}\hat{D}(\bar{\beta},\eta_0) \Vert=O_p(1)$. We thus have the following preliminary results,
\begin{align}
\label{eq:Jdiff}
    (a)&\quad \Vert \frac{\sqrt{n}}{\mu_n}\{\hat D(\hat\beta,\hat\eta)-\hat D(\bar{\beta},\eta_0)\}\Vert=o_p(1);\\
\label{eq:Jnorm}
    (b)&\quad \Vert \frac{\sqrt{n}}{\mu_n} \hat{D}(\hat\beta,\eta_0)\Vert=O_p(1),\quad \Vert \frac{\sqrt{n}}{\mu_n} \hat{D}(\bar{\beta},\eta_0)\Vert=O_p(1);\\
\label{eq:estomega}
    (c)&\quad \Vert \hat{\Omega}^{-1}(\hat{\beta},\hat{\eta})\Vert=O_p(1),\quad  \Vert \hat{\Omega}^{-1}(\bar{\beta},\eta_0)\Vert=O_p(1).
\end{align}
 By Lemma \ref{lemma:deQ} and Lemma A13 in \cite{Newey:2009aa}, and the continuous mapping theorem, 
 \begin{equation}
 \label{equ::de2Qconv}
     \left\{n\mu_n^{-2}\frac{\partial^2\hat{Q}(\hat{\beta},\hat{\eta})}{\partial\beta^2}\right\}^{-2}\overset{p}{\rightarrow}\frac{1}{\tau^2},
 \end{equation} so that $A_1=o_p(1)$. Moreover, by the Cauchy-Schwarz inequality,
\begin{align*}
    &|\mu_n^{-1}n^{1/2}[\hat{D}-\hat{D}(\bar{\beta},\eta_0)]^\prime\hat{\Omega}^{-1}(\hat{\beta},\hat{\eta})[\mu_n^{-1}n^{1/2}\hat{D}]|\\
    \leq& \Vert \frac{\sqrt{n}}{\mu_n}[\hat{D}-\hat{D}(\bar{\beta},\eta_0)]\Vert\Vert \hat{\Omega}^{-1}(\hat{\beta},\hat{\eta})\Vert\Vert \frac{\sqrt{n}}{\mu_n}\hat{D}\Vert\\
    =&o_p(1),
\end{align*}
where the equality holds by Assumption 2, Lemma \ref{lemma:cov} and (\ref{eq:Jdiff})-(\ref{eq:estomega}), so that we could have $A_2=o_p(1)$. We can show $A_4=o_p(1)$ in a similar way.  For $A_3$, 
\begin{align*}
    &\Vert \hat{\Omega}^{-1}(\hat{\beta},\hat{\eta})-\hat{\Omega}^{-1}(\bar{\beta},\eta_0)\Vert\\
    =&\Vert \hat\Omega^{-1}(\bar{\beta},\eta_0)\{\hat\Omega(\bar{\beta},\eta_0)-\hat\Omega(\hat\beta,\hat\eta)\}\hat\Omega^{-1}(\hat\beta,\hat\eta)\Vert\\
    \leq& C\Vert \hat\Omega(\bar{\beta},\eta_0)-\hat\Omega(\hat\beta,\hat\eta)\Vert\quad w.p.1\\
    \leq& C(\Vert \hat\Omega(\bar{\beta},\eta_0)-\hat\Omega(\beta_0,\eta_0)\Vert+\Vert \hat\Omega(\beta_0,\eta_0) - \hat\Omega(\beta_0,\hat\eta)\Vert)+\Vert \hat\Omega(\beta_0,\hat\eta) - \hat\Omega(\hat\beta,\hat\eta)\Vert\\
    = & \Vert \frac{1}{n} \sum_{i=1}^n g_i(\bar{\beta},\eta_0)g_i(\bar{\beta},\eta_0)^\prime - \frac{1}{n} \sum_{i=1}^n g_i(\beta_0,\eta_0)g_i(\beta_0,\eta_0)^\prime\Vert + o_p(1)\\
    +& \Vert \frac{1}{n} \sum_{i=1}^n g_i(\hat{\beta},\hat\eta)g_i(\hat{\beta},\hat\eta)^\prime - \frac{1}{n} \sum_{i=1}^n g_i(\beta_0,\hat\eta)g_i(\beta_0,\hat\eta)^\prime\Vert
    \\
    \leq&C\biggr\{|\bar{\beta}-\beta_0|\biggr\Vert n^{-1}\underset{i=1}{\overset{n}{\sum}}g_i(\beta_0,\eta_0)G_i(\eta_0)\biggr\Vert+ |\bar{\beta}-\beta_0|^2\biggr\Vert n^{-1} \underset{i=1}{\overset{n}{\sum}}G_i(\eta_0)G^\prime_i(\eta_0)\biggr\Vert\\
    &+|\hat{\beta}-\beta_0|\biggr\Vert n^{-1}\underset{i=1}{\overset{n}{\sum}}g_i(\beta_0,\hat\eta)G_i(\hat\eta)\biggr\Vert+|\hat{\beta}-\beta_0|^2\biggr\Vert n^{-1}\underset{i=1}{\overset{n}{\sum}}G_i(\hat\eta)G^\prime_i(\hat\eta)\biggr\Vert+o_p(1)\biggr\}\\ 
    =&o_p(1),
\end{align*}
where the first inequality holds by (\ref{eq:estomega}), the second inequality by the triangle inequality, the second equality  by Lemma \ref{lemma:cov}, and the last equality  by Lemmas \ref{lemma:omega}, \ref{lemma:cov} and the fact that both estimators of $\beta_0$ are consistent. In conclusion, we have that $\mu_n^2\hat{V}/n$ is asymptotically equivalent to $\mu_n^2\hat{V}(\bar{\beta},\eta_0)/n$. 
The consistency of the variance estimator then follows from Lemma A14 in \cite{Newey:2009aa}. To prove (i), first we have
\begin{align*}
    &\frac{n(\hat\beta-\beta_0)^2}{\hat V}\\
    =&\frac{\mu_n^2(\hat\beta-\beta_0)^2}{\mu_n^2\hat V/n}\\
    =&\biggr\{\frac{\mu_n(\hat\beta-\beta_0)}{\sqrt{\mu_n^2\hat V/n}}\biggr\}^2,
\end{align*}
then the result holds by $\mu_n^2\hat{V}/n\overset{p}{\rightarrow}V$, Slutsky's theorem and the continuous mapping theorem. For the LM statistics in part (ii), under the null, we have
$$\hat{K}
    =\biggr\{n\mu_n^{-1}\frac{\partial\hat Q(\beta_0)}{\partial\beta}\biggr\}\biggr\{ \frac{n\hat{D}(\beta_0)^\prime\hat{\Omega}(\beta_0)\hat D(\beta_0)}{\mu_n^2}\biggr\}^{-1}\biggr\{n\mu_n^{-1}\frac{\partial\hat Q(\beta_0)}{\partial\beta} \biggr\}.
$$
By the proof of consistency in the variance estimator and changing $\hat{\beta}$ to $\beta_0$, we have 
$$\biggr\{ \frac{n\hat{D}(\beta_0,\hat\eta)^\prime\hat{\Omega}(\beta_0,\hat\eta)\hat D(\beta_0,\hat\eta)}{\mu_n^2}\biggr\}^{-1}\overset{p}{\rightarrow}(\tau+\sigma^2)^{-1}. $$
Moreover, following the Taylor expansion and (\ref{equ::de2Qconv}), we have
$$ n\mu_n^{-1}\frac{\partial\hat Q(\beta_0,\hat\eta)}{\partial\beta}=\mu_n\tau(\hat\beta-\beta_0)+o_p(1).$$
Thus by the compactness of $\mathcal{B}$, $\hat{K}$ can be written as
\begin{align*}
    \hat{K}=&(\hat\beta-\beta_0)\mu_n\tau[(\tau+\sigma^2)^{-1}+o_p(1)]\tau\mu_n(\hat\beta-\beta_0)\\
    =&\biggr\{\frac{\mu_n(\hat\beta-\beta_0)}{\sqrt{V}}\biggr\}^2+o_p(1)\\
    =&\hat T+o_p(1).  
\end{align*}
Finally part (iii) follows from \citet[Theorem 3]{ye2024genius}.
\end{proof}
\clearpage
\bibliographystyle{agsm}
\bibliography{refs}

@misc{zhang2025debiasedcontinuousupdatinggmm,
      title={Debiased Continuous Updating GMM with Many Weak Instruments}, 
      author={Di Zhang and Baoluo Sun},
      year={2025},
      eprint={2504.18107},
      archivePrefix={arXiv}
}

@article{zeng2024causal,
  title={Causal inference with high-dimensional discrete covariates},
  author={Zeng, Zhenghao and Balakrishnan, Sivaraman and Han, Yanjun and Kennedy, Edward H},
  journal={arXiv preprint arXiv:2405.00118},
  year={2024}
}

@article{tibshirani1996regression,
  title={Regression shrinkage and selection via the lasso},
  author={Tibshirani, Robert},
  journal={Journal of the Royal Statistical Society Series B: Statistical Methodology},
  volume={58},
  number={1},
  pages={267--288},
  year={1996},
  publisher={Oxford University Press}
}

@article{dukes2020doubly,
  title={Doubly robust tests of exposure effects under high-dimensional confounding},
  author={Dukes, Oliver and Avagyan, Vahe and Vansteelandt, Stijn},
  journal={Biometrics},
  volume={76},
  number={4},
  pages={1190--1200},
  year={2020},
  publisher={Oxford University Press}
}

@article{caruana1997multitask,
  title={Multitask learning},
  author={Caruana, Rich},
  journal={Machine learning},
  volume={28},
  number={1},
  pages={41--75},
  year={1997},
  publisher={Springer}
}

@article{behdin2025multi,
  title={Multi-task learning for sparsity pattern heterogeneity: statistical and computational perspectives},
  author={Behdin, Kayhan and Loewinger, Gabriel and Kishida, Kenneth T and Parmigiani, Giovanni and Mazumder, Rahul},
  journal={Journal of the Royal Statistical Society Series B: Statistical Methodology},
  pages={qkaf076},
  year={2025},
  publisher={Oxford University Press UK}
}

@article{ning2020robust,
  title={Robust estimation of causal effects via a high-dimensional covariate balancing propensity score},
  author={Ning, Yang and Sida, Peng and Imai, Kosuke},
  journal={Biometrika},
  volume={107},
  number={3},
  pages={533--554},
  year={2020},
  publisher={Oxford University Press}
}

@article{10.1214/19-AOS1824,
author = {Tan, Zhiqiang},
title = {{Model-assisted inference for treatment effects using regularized calibrated estimation with high-dimensional data}},
volume = {48},
journal = {The Annals of Statistics},
number = {2},
publisher = {Institute of Mathematical Statistics},
pages = {811 -- 837},
year = {2020}
}

@article{tan2020regularized,
  title={Regularized calibrated estimation of propensity scores with model misspecification and high-dimensional data},
  author={Tan, Zhiqiang},
  journal={Biometrika},
  volume={107},
  number={1},
  pages={137--158},
  year={2020},
  publisher={Oxford University Press}
}

@article{dukes2021inference,
  title={Inference for treatment effect parameters in potentially misspecified high-dimensional models},
  author={Dukes, Oliver and Vansteelandt, Stijn},
  journal={Biometrika},
  volume={108},
  number={2},
  pages={321--334},
  year={2021},
  publisher={Oxford University Press}
}

@article{avagyan2022high,
  title={High-dimensional inference for the average treatment effect under model misspecification using penalized bias-reduced double-robust estimation},
  author={Avagyan, Vahe and Vansteelandt, Stijn},
  journal={Biostatistics \& Epidemiology},
  volume={6},
  number={2},
  pages={221--238},
  year={2022},
  publisher={Taylor \& Francis}
}

@article{athey2018approximate,
  title={Approximate residual balancing: debiased inference of average treatment effects in high dimensions},
  author={Athey, Susan and Imbens, Guido W and Wager, Stefan},
  journal={Journal of the Royal Statistical Society Series B: Statistical Methodology},
  volume={80},
  number={4},
  pages={597--623},
  year={2018},
  publisher={Oxford University Press}
}

@article{tang2023ultra,
  title={Ultra-high dimensional variable selection for doubly robust causal inference},
  author={Tang, Dingke and Kong, Dehan and Pan, Wenliang and Wang, Linbo},
  journal={Biometrics},
  volume={79},
  number={2},
  pages={903--914},
  year={2023},
  publisher={Wiley Online Library}
}

@article{farrell2015robust,
  title={Robust inference on average treatment effects with possibly more covariates than observations},
  author={Farrell, Max H},
  journal={Journal of Econometrics},
  volume={189},
  number={1},
  pages={1--23},
  year={2015},
  publisher={Elsevier}
}

@article{mikusheva2022inference,
  title={Inference with many weak instruments},
  author={Mikusheva, Anna and Sun, Liyang},
  journal={The Review of Economic Studies},
  volume={89},
  number={5},
  pages={2663--2686},
  year={2022},
  publisher={Oxford University Press}
}

@TechReport{repec:pri:econom:2018-16,
type={Working Papers},
institution={Princeton University. Economics Department.},
author={Kirill S. Evdokimov and Michal Koles{\'a}r},
title={Inference in Instrumental Variable Regression Analysis with Heterogeneous Treatment Effects},
year={2018}
}

@article{https://doi.org/10.1111/rssb.12026,
author = {Zhang, Cun-Hui and Zhang, Stephanie S.},
title = {Confidence intervals for low dimensional parameters in high dimensional linear models},
journal = {Journal of the Royal Statistical Society: Series B (Statistical Methodology)},
volume = {76},
number = {1},
pages = {217-242},
year = {2014}
}

@article{bradic2019sparsity,
  title={Sparsity double robust inference of average treatment effects},
  author={Bradic, Jelena and Wager, Stefan and Zhu, Yinchu},
  journal={arXiv preprint arXiv:1905.00744},
  year={2019}
}

@article{bradic2019minimax,
  title={Minimax semiparametric learning with approximate sparsity},
  author={Bradic, Jelena and Chernozhukov, Victor and Newey, Whitney K and Zhu, Yinchu},
  journal={arXiv preprint arXiv:1912.12213},
  year={2019}
}

@article{hirshberg2021augmented,
  title={Augmented minimax linear estimation},
  author={Hirshberg, David A and Wager, Stefan},
  journal={The Annals of Statistics},
  volume={49},
  number={6},
  pages={3206--3227},
  year={2021},
  publisher={Institute of Mathematical Statistics}
}

@article{donald2001choosing,
  title={Choosing the number of instruments},
  author={Donald, Stephen G and Newey, Whitney K},
  journal={Econometrica},
  volume={69},
  number={5},
  pages={1161--1191},
  year={2001},
  publisher={Wiley Online Library}
}

@article{10.1214/14-STS493,
author = {Stijn Vansteelandt and Marshall Joffe},
title = {{Structural Nested Models and G-estimation: The Partially Realized Promise}},
volume = {29},
journal = {Statistical Science},
number = {4},
publisher = {Institute of Mathematical Statistics},
pages = {707 -- 731},
year = {2014}
}

@article{wang2025gmm,
  title={{GMM} with Many Weak Moment Conditions and Nuisance Parameters: {General} Theory and Applications to Causal Inference},
  author={Wang, Rui and Chan, Kwun Chuen Gary and Ye, Ting},
  journal={arXiv preprint arXiv:2505.07295},
  year={2025}
}

@article{carrasco2015regularized,
  title={Regularized LIML for many instruments},
  author={Carrasco, Marine and Tchuente, Guy},
  journal={Journal of Econometrics},
  volume={186},
  number={2},
  pages={427--442},
  year={2015},
  publisher={Elsevier}
}

@article{angrist1991does,
  title={Does compulsory school attendance affect schooling and earnings?},
  author={Angrist, Joshua D and Krueger, Alan B},
  journal={The Quarterly Journal of Economics},
  volume={106},
  number={4},
  pages={979--1014},
  year={1991},
  publisher={MIT Press}
}

@article{pfanzagl1982lecture,
  title={Lecture notes in statistics},
  author={Pfanzagl, Johann},
  journal={Contributions to a general asymptotic statistical theory},
  volume={13},
  pages={11--15},
  year={1982},
  publisher={Springer}
}

@book{van2000asymptotic,
  title={Asymptotic statistics},
  author={Van der Vaart, Aad W},
  volume={3},
  year={2000},
  publisher={Cambridge university press}
}

@article{newey1994asymptotic,
  title={The asymptotic variance of semiparametric estimators},
  author={Newey, Whitney K},
  journal={Econometrica},
  pages={1349--1382},
  year={1994},
  publisher={JSTOR}
}

@article{andrews1994asymptotics,
  title={Asymptotics for semiparametric econometric models via stochastic equicontinuity},
  author={Andrews, Donald WK},
  journal={Econometrica},
  pages={43--72},
  year={1994},
  publisher={JSTOR}
}

@phdthesis{ayyagari2010applications,
  title={Applications of influence functions to semiparametric regression models},
  author={Ayyagari, Rajeev},
  year={2010},
  school={Harvard University}
}

@article{newey2018cross,
  title={Cross-fitting and fast remainder rates for semiparametric estimation},
  author={Newey, Whitney K and Robins, James R},
  journal={arXiv preprint arXiv:1801.09138},
  year={2018}
}

@article{chernozhukov2017double,
  title={Double/debiased/neyman machine learning of treatment effects},
  author={Chernozhukov, Victor and Chetverikov, Denis and Demirer, Mert and Duflo, Esther and Hansen, Christian and Newey, Whitney},
  journal={American Economic Review},
  volume={107},
  number={5},
  pages={261--265},
  year={2017},
  publisher={American Economic Association 2014 Broadway, Suite 305, Nashville, TN 37203}
}

@article{scharfstein1999adjusting,
  title={Adjusting for nonignorable drop-out using semiparametric nonresponse models},
  author={Scharfstein, Daniel O and Rotnitzky, Andrea and Robins, James M},
  journal={Journal of the American Statistical Association},
  volume={94},
  number={448},
  pages={1096--1120},
  year={1999},
  publisher={Taylor \& Francis}
}

@article{guo2022robustness,
  title={Robustness against weak or invalid instruments: Exploring nonlinear treatment models with machine learning},
  author={Guo, Zijian and Zheng, Mengchu and B{\"u}hlmann, Peter},
  journal={arXiv preprint arXiv:2203.12808},
  year={2022}
}

@inproceedings{hasminskii1979nonparametric,
  title={On the nonparametric estimation of functionals},
  author={Hasminskii, Rafail Z and Ibragimov, Ildar A},
  booktitle={{Proceedings of the Second Prague Symposium on Asymptotic Statistics}},
  volume={473},
  pages={474--482},
  year={1979},
  organization={North-Holland Amsterdam}
}

@article{bickel1982adaptive,
  title={On adaptive estimation},
  author={Bickel, Peter J},
  journal={The Annals of Statistics},
  pages={647--671},
  year={1982},
  publisher={JSTOR}
}

@article{chen2020mostly,
  title={{Mostly harmless machine learning: learning optimal instruments in linear IV models}},
  author={Chen, Jiafeng and Chen, Daniel L and Lewis, Greg},
  journal={arXiv preprint arXiv:2011.06158},
  year={2020}
}

@article{liu2020deep,
  title={On deep instrumental variables estimate},
  author={Liu, Ruiqi and Shang, Zuofeng and Cheng, Guang},
  journal={arXiv preprint arXiv:2004.14954},
  year={2020}
}

@article{belloni2012sparse,
  title={Sparse models and methods for optimal instruments with an application to eminent domain},
  author={Belloni, Alexandre and Chen, Daniel and Chernozhukov, Victor and Hansen, Christian},
  journal={Econometrica},
  volume={80},
  number={6},
  pages={2369--2429},
  year={2012},
  publisher={Wiley Online Library}
}

@article{van2006targeted,
  title={Targeted maximum likelihood learning},
  author={{Van Der Laan}, Mark J and Rubin, Daniel},
  journal={The International Journal of Biostatistics},
  volume={2},
  number={1},
  year={2006},
  publisher={De Gruyter}
}

@Inbook{Zheng2011,
author="Zheng, Wenjing
and {van der Laan}, Mark J.",
title="Cross-Validated Targeted Minimum-Loss-Based Estimation",
bookTitle="Targeted Learning: Causal Inference for Observational and Experimental Data",
year="2011",
publisher="Springer New York",
address="New York, NY",
pages="459--474"
}

@incollection{neyman1,
  title={Optimal asymptotic tests of composite statistical hypotheses},
  author={Jerzy Neyman},
  booktitle={Probability and Statistics},
  pages={416--44},
  year={1959},
  publisher={Wiley}
}

@article{neyman2,
  title={$c(\alpha)$ tests and their use},
  author={Jerzy Neyman},
  journal={Sankhya},
  year={1979},
  pages={1--21},
}

@book{van2011targeted,
  title={Targeted learning: causal inference for observational and experimental data},
  author={{Van der Laan}, Mark J and Rose, Sherri and others},
  volume={4},
  year={2011},
  publisher={Springer}
}

@article{hansen1996finite,
  title={Finite-sample properties of some alternative {GMM} estimators},
  author={Hansen, Lars Peter and Heaton, John and Yaron, Amir},
  journal={Journal of Business \& Economic Statistics},
  volume={14},
  number={3},
  pages={262--280},
  year={1996},
  publisher={Taylor \& Francis}
}

@article{han2006gmm,
  title={GMM with many moment conditions},
  author={Han, Chirok and Phillips, Peter CB},
  journal={Econometrica},
  volume={74},
  number={1},
  pages={147--192},
  year={2006},
  publisher={Wiley Online Library}
}

@article{diaz2020machine,
  title={Machine learning in the estimation of causal effects: targeted minimum loss-based estimation and double/debiased machine learning},
  author={D{\'\i}az, Iv{\'a}n},
  journal={Biostatistics},
  volume={21},
  number={2},
  pages={353--358},
  year={2020},
  publisher={Oxford University Press}
}

@article{hines2022demystifying,
  title={Demystifying statistical learning based on efficient influence functions},
  author={Hines, Oliver and Dukes, Oliver and Diaz-Ordaz, Karla and Vansteelandt, Stijn},
  journal={The American Statistician},
  volume={76},
  number={3},
  pages={292--304},
  year={2022},
  publisher={Taylor \& Francis}
}

@article{shi2020multiply,
  title={Multiply robust causal inference with double-negative control adjustment for categorical unmeasured confounding},
  author={Shi, Xu and Miao, Wang and Nelson, Jennifer C and {Tchetgen Tchetgen}, Eric J},
  journal={Journal of the Royal Statistical Society Series B: Statistical Methodology},
  volume={82},
  number={2},
  pages={521--540},
  year={2020},
  publisher={Oxford University Press}
}

@article{cui2024semiparametric,
  title={Semiparametric proximal causal inference},
  author={Cui, Yifan and Pu, Hongming and Shi, Xu and Miao, Wang and {Tchetgen Tchetgen}, Eric},
  journal={Journal of the American Statistical Association},
  volume={119},
  number={546},
  pages={1348--1359},
  year={2024},
  publisher={Taylor \& Francis}
}

@article{scheidegger2025inference,
  title={Inference for Heterogeneous Treatment Effects with Efficient Instruments and Machine Learning},
  author={Scheidegger, Cyrill and Guo, Zijian and B{\"u}hlmann, Peter},
  journal={arXiv preprint arXiv:2503.03530},
  year={2025}
}

@article{miao2017invited,
  title={Invited commentary: bias attenuation and identification of causal effects with multiple negative controls},
  author={Miao, Wang and {Tchetgen Tchetgen}, Eric},
  journal={American journal of epidemiology},
  volume={185},
  number={10},
  pages={950--953},
  year={2017},
  publisher={Oxford University Press}
}

@article{tchetgen2024introduction,
  title={An introduction to proximal causal inference},
  author={{Tchetgen Tchetgen}, Eric J and Ying, Andrew and Cui, Yifan and Shi, Xu and Miao, Wang},
  journal={Statistical Science},
  volume={39},
  number={3},
  pages={375--390},
  year={2024},
  publisher={Institute of Mathematical Statistics}
}

@article{davies2015many,
  title={The many weak instruments problem and {Mendelian} randomization},
  author={Davies, Neil M and von Hinke Kessler Scholder, Stephanie and Farbmacher, Helmut and Burgess, Stephen and Windmeijer, Frank and Smith, George Davey},
  journal={Stat. Med.},
  volume={34},
  number={3},
  pages={454--468},
  year={2015},
  publisher={Wiley Online Library}
}

@article{hansen2008estimation,
  title={Estimation with many instrumental variables},
  author={Hansen, Christian and Hausman, Jerry and Newey, Whitney},
  journal={Journal of Business \& Economic Statistics},
  volume={26},
  number={4},
  pages={398--422},
  year={2008},
  publisher={Taylor \& Francis}
}

@article{kleibergen2005testing,
  title={Testing parameters in GMM without assuming that they are identified},
  author={Kleibergen, Frank},
  journal={Econometrica},
  volume={73},
  number={4},
  pages={1103--1123},
  year={2005},
  publisher={Wiley Online Library}
}

@article{guggenberger2005generalized,
  title={Generalized empirical likelihood estimators and tests under partial, weak, and strong identification},
  author={Guggenberger, Patrik and Smith, Richard J},
  journal={Econometric Theory},
  volume={21},
  number={4},
  pages={667--709},
  year={2005},
  publisher={Cambridge University Press}
}

@article{bound1995problems,
  title={Problems with instrumental variables estimation when the correlation between the instruments and the endogenous explanatory variable is weak},
  author={Bound, John and Jaeger, David A and Baker, Regina M},
	Journal = {J. Am. Statist. Assoc.},
  volume={90},
  number={430},
  pages={443--450},
  year={1995},
  publisher={Taylor \& Francis}
}

@article{chernozhukov2022locally,
  title={Locally robust semiparametric estimation},
  author={Chernozhukov, Victor and Escanciano, Juan Carlos and Ichimura, Hidehiko and Newey, Whitney K and Robins, James M},
  journal={Econometrica},
  volume={90},
  number={4},
  pages={1501--1535},
  year={2022},
  publisher={Wiley Online Library}
}

@article{chao2005consistent,
  title={Consistent estimation with a large number of weak instruments},
  author={Chao, John C and Swanson, Norman R},
  journal={Econometrica},
  volume={73},
  number={5},
  pages={1673--1692},
  year={2005},
  publisher={Wiley Online Library}
}

@article{Staiger:1997aa,
	Author = {Staiger, Douglas and Stock, James H.},
	C1 = {Full publication date: May, 1997},
	Db = {JSTOR},
	Isbn = {00129682, 14680262},
	Journal = {Econometrica},
	Number = {3},
	Pages = {557--586},
	Publisher = {{$[$}Wiley, Econometric Society{$]$}},
	Title = {Instrumental Variables Regression with Weak Instruments},
	Ty = {JOUR},
	Volume = {65},
	Year = {1997}
}

@article{stock2002survey,
  title={A survey of weak instruments and weak identification in generalized method of moments},
  author={Stock, James H and Wright, Jonathan H and Yogo, Motohiro},
  journal={J. Bus. Econ. Statist.},
  volume={20},
  number={4},
  pages={518--529},
  year={2002},
  publisher={Taylor \& Francis}
}

@article{bekker1994alternative,
  title={Alternative approximations to the distributions of instrumental variable estimators},
  author={Bekker, Paul A},
  journal={Econometrica},
  pages={657--681},
  year={1994},
  publisher={JSTOR}
}

@article{Chao:2005aa,
	Author = {Chao, John C. and Swanson, Norman R.},
	Booktitle = {Econometrica},
	Da = {2005/09/01},
	Date = {2005/09/01},
	Isbn = {0012-9682},
	Journal = {Econometrica},
	Journal1 = {Econometrica},
	Number = {5},
	Pages = {1673--1692},
	Publisher = {John Wiley \& Sons, Ltd (10.1111)},
	Title = {Consistent Estimation with a Large Number of Weak Instruments},
	Ty = {JOUR},
	Volume = {73},
	Year = {2005},
	Year1 = {2005}}

@article{Newey:2009aa,
	Author = {Newey, Whitney K. and Windmeijer, Frank},
	Booktitle = {Econometrica},
	Da = {2009/05/01},
	Date = {2009/05/01},
	Isbn = {0012-9682},
	Journal = {Econometrica},
	Journal1 = {Econometrica},
	Number = {3},
	Pages = {687-719},
	Publisher = {John Wiley \& Sons, Ltd},
	Title = {Generalized Method of Moments With Many Weak Moment Conditions},
	Ty = {JOUR},
	Volume = {77},
	Year = {2009},
	Year1 = {2009}}

@article{ye2024genius,
  title={{GENIUS-MAWII}: For robust {Mendelian} randomization with many weak invalid instruments},
  author={Ye, Ting and Liu, Zhonghua and Sun, Baoluo and Tchetgen Tchetgen, Eric},
	Journal = {J. R. Stat. Soc. {\normalfont B}},
  pages={qkae024},
  year={2024},
  publisher={Oxford University Press US}
}

@article{robins1994correcting,
  title={Correcting for non-compliance in randomized trials using structural nested mean models},
  author={Robins, James M},
  journal={Commun. Statist. {\normalfont A}},
  volume={23},
  number={8},
  pages={2379--2412},
  year={1994},
  publisher={Taylor \& Francis}
}

@article{smucler2019unifying,
  title={A unifying approach for doubly-robust $\ell_1 $ regularized estimation of causal contrasts},
  author={Smucler, Ezequiel and Rotnitzky, Andrea and Robins, James M},
  journal={arXiv preprint arXiv:1904.03737},
  year={2019}
}

@article{lounici2009taking,
  title={Taking advantage of sparsity in multi-task learning},
  author={Lounici, Karim and Pontil, Massimiliano and Tsybakov, Alexandre B and Van De Geer, Sara},
  journal={arXiv preprint arXiv:0903.1468},
  year={2009}
}

@article{sargan1958estimation,
  title={The estimation of economic relationships using instrumental variables},
  author={Sargan, John D},
  journal={Econometrica},
  pages={393--415},
    volume={26},
  number={3},
  year={1958},
  publisher={JSTOR}
}

@article{Hansen:1982aa,
	Author = {Hansen, Lars Peter},
	C1 = {Full publication date: Jul., 1982},
	Db = {JSTOR},
	Doi = {10.2307/1912775},
	Isbn = {00129682, 14680262},
	Journal = {Econometrica},
	Number = {4},
	Pages = {1029--1054},
	Publisher = {{$[$}Wiley, Econometric Society{$]$}},
	Title = {Large Sample Properties of Generalized Method of Moments Estimators},
	Ty = {JOUR},
	Volume = {50},
	Year = {1982}}

@article{rotnitzky2021characterization,
  title={Characterization of parameters with a mixed bias property},
  author={Rotnitzky, Andrea and Smucler, Ezequiel and Robins, James M},
  journal={Biometrika},
  volume={108},
  number={1},
  pages={231--238},
  year={2021},
  publisher={Oxford University Press}
}

@article{okui2012doubly,
  title={Doubly robust instrumental variable regression},
  author={Okui, Ryo and Small, Dylan S and Tan, Zhiqiang and Robins, James M},
  journal={Statist. Sinica},
  pages={173--205},
  year={2012},
  publisher={JSTOR}
}

@article{Chernozhukov2018ddml,
	author = {Chernozhukov, Victor and Chetverikov, Denis and Demirer, Mert and Duflo, Esther and Hansen, Christian and Newey, Whitney and Robins, James},
	title = "{Double/debiased machine learning for treatment and structural parameters}",
	journal = {Economet. J.},
	volume = {21},
	number = {1},
	pages = {C1-C68},
	year = {2018},
	month = {01},
	issn = {1368-4221},
	doi = {10.1111/ectj.12097},
	url = {https://doi.org/10.1111/ectj.12097},
	eprint = {https://academic.oup.com/ectj/article-pdf/21/1/C1/27684918/ectj00c1.pdf},
}

@article{tripathi1999matrix,
  title={A matrix extension of the Cauchy-Schwarz inequality},
  author={Tripathi, Gautam},
  journal={Economics Letters},
  volume={63},
  number={1},
  pages={1--3},
  year={1999},
  publisher={Elsevier}
}

@article{Yuan_2005, title={Model Selection and Estimation in Regression with Grouped Variables}, volume={68}, ISSN={1467-9868}, number={1}, journal={Journal of the Royal Statistical Society Series B: Statistical Methodology}, publisher={Oxford University Press (OUP)}, author={Yuan, Ming and Lin, Yi}, year={2005}, month=dec, pages={49–67} }

@article{Lounici_2011, title={Oracle inequalities and optimal inference under group sparsity}, volume={39}, ISSN={0090-5364}, DOI={10.1214/11-aos896}, number={4}, journal={The Annals of Statistics}, publisher={Institute of Mathematical Statistics}, author={Lounici, Karim and Pontil, Massimiliano and van de Geer, Sara and Tsybakov, Alexandre B.}, year={2011}, month=aug }

@article{negahban_unified_2012,
	title = {A {Unified} {Framework} for {High}-{Dimensional} {Analysis} of \${M}\$-{Estimators} with {Decomposable} {Regularizers}},
	volume = {27},
	issn = {0883-4237},
	doi = {10.1214/12-STS400},
	language = {en},
	number = {4},
	urldate = {2025-09-08},
	journal = {Statistical Science},
	author = {Negahban, Sahand N. and Ravikumar, Pradeep and Wainwright, Martin J. and Yu, Bin},
	month = nov,
	year = {2012},
}

@article{blazere_oracle_2014,
	title = {Oracle {Inequalities} for a {Group} {Lasso} {Procedure} {Applied} to {Generalized} {Linear} {Models} in {High} {Dimension}},
	volume = {60},
	issn = {1557-9654},
	doi = {10.1109/TIT.2014.2303121},
	language = {en-US},
	number = {4},
	urldate = {2025-09-09},
	journal = {IEEE Transactions on Information Theory},
	author = {Blazère, Mélanie and Loubes, Jean-Michel and Gamboa, Fabrice},
	month = apr,
	year = {2014},
	pages = {2303--2318},
}

@book{wainwright2019high,
  title={High-dimensional statistics: A non-asymptotic viewpoint},
  author={Wainwright, Martin J},
  volume={48},
  year={2019},
  publisher={Cambridge university press}
}

@article{mikusheva2024weak,
  title={Weak identification with many instruments},
  author={Mikusheva, Anna and Sun, Liyang},
  journal={The Econometrics Journal},
  volume={27},
  number={2},
  pages={C1--C28},
  year={2024},
  publisher={Oxford University Press}
}

@article{chernozhukov2015post,
  title={Post-selection and post-regularization inference in linear models with many controls and instruments},
  author={Chernozhukov, Victor and Hansen, Christian and Spindler, Martin},
  journal={American Economic Review},
  volume={105},
  number={5},
  pages={486--490},
  year={2015},
  publisher={American Economic Association 2014 Broadway, Suite 305, Nashville, TN 37203}
}

@article{chao2023jackknife,
  title={Jackknife estimation of a cluster-sample IV regression model with many weak instruments},
  author={Chao, John C and Swanson, Norman R and Woutersen, Tiemen},
  journal={Journal of Econometrics},
  volume={235},
  number={2},
  pages={1747--1769},
  year={2023},
  publisher={Elsevier}
}

@article{meier_group_2008,
    title = {The {Group} {Lasso} for {Logistic} {Regression}},
    volume = {70},
    copyright = {https://academic.oup.com/journals/pages/open\_access/funder\_policies/chorus/standard\_publication\_model},
    issn = {1369-7412, 1467-9868},
    url = {https://academic.oup.com/jrsssb/article/70/1/53/7109381},
    doi = {10.1111/j.1467-9868.2007.00627.x},
    language = {en},
    number = {1},
    urldate = {2025-10-24},
    journal = {Journal of the Royal Statistical Society Series B: Statistical Methodology},
    author = {Meier, Lukas and Van De Geer, Sara and Bühlmann, Peter},
    month = feb,
    year = {2008},
    pages = {53--71},
}

@article{van_de_geer_asymptotically_2014,
    title = {On asymptotically optimal confidence regions and tests for high-dimensional models},
    volume = {42},
    issn = {0090-5364},
    doi = {10.1214/14-AOS1221},
    language = {en},
    number = {3},
    urldate = {2025-10-24},
    journal = {The Annals of Statistics},
    author = {Van De Geer, Sara and Bühlmann, Peter and Ritov, Ya’acov and Dezeure, Ruben},
    year = {2014},
}

@article{cao_rmtl_2019,
    title = {{RMTL}: an {R} library for multi-task learning},
    volume = {35},
    issn = {1367-4803},
    shorttitle = {{RMTL}},
    url = {https://doi.org/10.1093/bioinformatics/bty831},
    doi = {10.1093/bioinformatics/bty831},
    number = {10},
    urldate = {2025-11-09},
    journal = {Bioinformatics},
    author = {Cao, Han and Zhou, Jiayu and Schwarz, Emanuel},
    month = may,
    year = {2019},
    pages = {1797--1798},
}

@article{chernozhukov2015valid,
  title={Valid post-selection and post-regularization inference: An elementary, general approach},
  author={Chernozhukov, Victor and Hansen, Christian and Spindler, Martin},
  journal={Annu. Rev. Econ.},
  volume={7},
  number={1},
  pages={649--688},
  year={2015},
  publisher={Annual Reviews}
}

@article{liu2023root,
  title={Root-n consistent semiparametric learning with high-dimensional nuisance functions under minimal sparsity},
  author={Liu, Lin and Wang, Xinbo and Wang, Yuhao},
  journal={arXiv preprint arXiv:2305.04174},
  year={2023}
}

@article{matsushita5367195cross,
  title={Cross-Fitted Instrumental Variable Estimation with Many Instruments and Many Included Exogenous Regressors},
  author={Matsushita, Yukitoshi and Otsu, Taisuke},
    year={2025},
  journal={Available at SSRN 5367195}
}

@misc{wainwright_sharp_2006,
    title = {Sharp thresholds for high-dimensional and noisy recovery of sparsity},
    url = {http://arxiv.org/abs/math/0605740},
    doi = {10.48550/arXiv.math/0605740},
    urldate = {2025-12-07},
    publisher = {arXiv},
    author = {Wainwright, Martin J.},
    month = may,
    year = {2006},
    note = {arXiv:math/0605740},
}

@article{cai_statistical_2023,
    title = {Statistical {Inference} for {High}-{Dimensional} {Generalized} {Linear} {Models} {With} {Binary} {Outcomes}},
    volume = {118},
    issn = {0162-1459},
    url = {https://doi.org/10.1080/01621459.2021.1990769},
    doi = {10.1080/01621459.2021.1990769},
    number = {542},
    urldate = {2025-09-19},
    journal = {Journal of the American Statistical Association},
    author = {Cai, T. Tony and Guo, Zijian and Ma, Rong},
    month = apr,
    year = {2023},
    pmid = {37366472},
    note = {Publisher: ASA Website
\_eprint: https://doi.org/10.1080/01621459.2021.1990769},
    pages = {1319--1332},
}

@article{koch_covariate_2018,
    title = {Covariate selection with group lasso and doubly robust estimation of causal effects},
    volume = {74},
    issn = {1541-0420},
    url = {https://onlinelibrary.wiley.com/doi/abs/10.1111/biom.12736},
    doi = {10.1111/biom.12736},
    language = {en},
    number = {1},
    urldate = {2025-12-07},
    journal = {Biometrics},
    author = {Koch, Brandon and Vock, David M. and Wolfson, Julian},
    year = {2018},
    note = {\_eprint: https://onlinelibrary.wiley.com/doi/pdf/10.1111/biom.12736},
    pages = {8--17},
}

\end{document}